\documentclass{jfm}

\usepackage{graphicx}
\usepackage{epstopdf,epsfig}
\usepackage{newtxtext}
\usepackage{newtxmath}
\usepackage{tikz}
\usetikzlibrary{shapes}
\usepackage{color}
\usepackage{natbib}
\usepackage{hyperref}
\usepackage{tabu}
\usepackage{longtable}
\usepackage[noabbrev]{cleveref}
\usepackage{ragged2e}

\newcommand{\RomanNumeralCaps}[1]
\linenumbers

\usepackage{bm}
\usepackage[final]{changes}

\newcommand{\ub}{{\bf u}}

\newcommand{\Ub}{{\bf U}}
\newcommand{\Xb}{{\bf X}}

\newcommand{\xb}{{\bf x}}

\newcommand{\fb}{{\bf f}}
\newcommand{\Fb}{{\bf F}}
\newcommand{\partialb}{{\boldsymbol \partial}}

\title{On the fully coupled dynamics of flexible fibres dispersed in modulated turbulence}

\author{Stefano Olivieri\aff{1} \corresp{\email{stefano.olivieri@oist.jp}},
Andrea Mazzino\aff{2,3}
\and Marco E. Rosti\aff{1}  \corresp{\email{marco.rosti@oist.jp}}}

\affiliation{
\aff{1} Complex Fluids and Flows Unit, Okinawa Institute of Science and Technology Graduate University, 1919-1 Tancha, Onna-son, Okinawa 904-0495, Japan\\
\aff{2} Department of Civil, Chemical and Environmental Engineering (DICCA), University of Genova, Via Montallegro 1, 16145, Genova, Italy \\
\aff{3} INFN, Genova Section, Via Montallegro 1, 16145, Genova, Italy
}

\begin{document}
\maketitle

\begin{abstract}
The present work investigates the mechanical behaviour of finite-size, elastic and inertial fibres freely moving in a homogeneous and isotropic turbulent flow at moderate Reynolds number. \added{Four-way coupled}, direct numerical simulations, based on a finite difference discretisation and the immersed boundary method, are performed to mutually couple the dynamics of fibres and fluid turbulence, allowing to account for the backreaction of the dispersed phase to the carrier flow. An extensive parametric study is carried out \added{in zero-gravity condition} over the characteristic properties of the suspension, i.e., fibre's linear density (from \added{iso-dense} to \added{denser-than-the-fluid} fibres), length (from short fibres comparable with the dissipative scale to long fibres comparable with the integral scale) and bending stiffness (from highly flexible to almost rigid fibres), as well as the concentration (from dilute to non-dilute suspensions). Results reveal the existence of a robust turbulence modulation mechanism that is primarily controlled by the mass fraction of the suspension (with only a minor influence of the fibre's bending stiffness), which is characterised in detail by means of a scale-by-scale analysis in Fourier space. Despite such alteration with respect to the single-phase case due to the non-negligible backreaction, fibres experience only two possible flapping states (previously identified in the very dilute condition) while being transported and deformed by the flow. Additionally, we show that the maximum curvature obeys to different scaling laws that can be derived from the fibre dynamical equation. Finally, we explore the \added{clustering} and \added{preferential} alignment of fibres within the flow, highlighting the peculiar role of inertia and elasticity.
\end{abstract}

\begin{keywords}
%Authors should not enter keywords on the manuscript, as \thispagestyle{•}ese must be chosen by the author during the online submission process and will then be added during the typesetting process (see \href{https://www.cambridge.org/core/journals/journal-of-fluid-mechanics/information/list-of-keywords}{Keyword PDF} for the full list).  Other classifications will be added at the same time.
\end{keywords}

{\bf MSC Codes }  {\it(Optional)} Please enter your MSC Codes here

\section{Introduction}
Suspensions of fibre-like objects dispersed in turbulent flows are encountered in a variety of natural and engineering problems, e.g., microplastics or non-motile microorganisms dispersal in aquatic environments, pulp production in papermaking, and other ecological or industrial processes~\citep{lundell2011fluid,guasto2012fluid,duroure2019review}. Several topics appear of particular relevance nowadays, such as the quantitative understanding of the fragmentation process of marine litter in the oceans~\citep{cozar2014plastic,filella2015questions,brouzet2021laboratory}. Other examples concern the locomotion of bacteria or planktonic organisms~\citep{son2013bacteria,ardekani2017sedimentation,michalec2017zooplankton}, the formation of algae aggregates~\citep{verhille2017structure}, and the manufacturing of paper and composite materials~\citep{parsheh2005orientation,mortensen2008dynamics,marchioli2010orientation}.

From a fundamental perspective, understanding the most salient features in this specific kind of multiphase flow represents a topic of active research. Compared to spherical particles, additional complexity is given by the anisotropic character of the dispersed objects, reflecting into peculiar phenomena such as the preferential alignment of the fibre's orientation with characteristic quantities of the flow, i.e., vorticity or strain rate principal directions~\citep{parsa2012rotation,voth2017anisotropic}. Focusing on rigid fibres (i.e., rods), numerous efforts have been devoted to describe in detail how such alignment, and consequently the fibre's rotation rate, are qualitatively and quantitatively conditioned by the fibre's length (compared to the characteristic lengthscales of the turbulent flow, ranging from sub-Kolmogorov to inertial subrange)~\citep{ni2014alignment,ni2015measurements,pujara2019scale,pujara2021shape,oehmke2021spinning} and inertia (expressed by means of a representative Stokes number)~\citep{bounoua2018tumbling,kuperman2019inertial}.

In many applications, however, fibres are not rigid but flexible, and may experience large deformations under the action of the flow, resulting in non-trivial dynamics already when immersed in simple laminar flows~\citep{duroure2019review,zuk2021universal}. While a relatively extensive research has focused on suspensions of rigid fibres, fewer studies are available on the case of flexible fibres in turbulence. Besides the relevance in the framework of particle-laden and multiphase flows, the latter represents an intriguing problem in the context of fluid-structure interaction (FSI), started to be investigated only quite recently~\citep{brouzet2014flexible,rosti2018flexible}. From the theoretical and numerical viewpoint, fibres are typically assumed to be inextensible, so that the attention is primarily devoted to the bending deformation~\citep{allende2020dynamics}, although some contributions focused on extensible objects as well, mostly related to suspensions of polymers~\citep{picardo2018preferential,picardo2020dynamics,vincenzi2021polymer}. The mechanical behaviour of flexible fibres in turbulence has been characterised with specific interest towards the buckling of short fibres, i.e., having sub-Kolmogorov length~\citep{allende2018stretching}, and the spatial conformation and deformation of finite-size fibres, i.e., with length within the inertial subrange or comparable to the integral lengthscale\added{~\citep{brouzet2014flexible,gay2018characterisation,sulaiman2019numerical,dotto2019orientation,dotto2020deformation,picardo2020dynamics}};
further recent developments include the aforementioned modelling of fragmentation processes~\citep{allende2020dynamics,brouzet2021laboratory,vincenzi2021polymer}.

From a relatively complementary perspective, recent studies focused on the possibility of exploiting fibre-like objects to measure the properties of the turbulent carrier flow, both in the case of flexible~\citep{rosti2018flexible,rosti2019flowing} and rigid fibres~\citep{brizzolara2021fibre}. For flexible fibres,~\citet{rosti2018flexible,rosti2019flowing} unravelled the existence of different dynamical states that can be predicted by comparing the characteristic timescales involved in the problem. In particular, it was shown that in certain regimes the fibres behave as a proxy of turbulent eddies of comparable size, therefore enabling the measurement of two-point flow statistics based on the longitudinal velocity differences by simply tracking the motion of the fibre (or, more practically, only the two fibre's ends). More recently,~\citet{olivieri2021universal} extended such findings to the case of a non-dilute suspension where the flow is substantially altered by the presence of the dispersed phase, showing that the same qualitative scenario is retained (although in the latter situation fibres do not measure the unperturbed flow anymore, but the fibre modulated one). In the case of rigid fibres, similar outcomes can be obtained when considering negligible inertia (i.e., vanishing Stokes number) and focusing on the transverse (instead of longitudinal) velocity differences: such findings have been recently corroborated in the laboratory framework by \citet{brizzolara2021fibre}, leading to the development of a novel experimental technique, named ``Fibre Tracking Velocimetry'', able to measure the properties of turbulence at a fixed lengthscale (i.e., the length of the fibre). Remarkably, exploiting this concept intrinsically overcomes the well-known issue of relative dispersion experienced with more traditional approaches based on tracer particles.  Moreover, it was demonstrated that for sufficiently short fibres the proposed technique leads to the accurate measurement of the turbulent energy dissipation rate. 

The majority of studies focusing on the dynamics of dispersed particles in turbulence are typically based on assuming that the suspension is dilute enough, so that the backreaction of the dispersed phase to the carrier flow can be safely ignored, and therefore dramatically simplifying the modelling of the problem by exploiting a \textit{one-way} coupling approach (i.e., the fluid flow is not affected by the presence of the dispersed objects). However,  one needs to consider the \textit{two-way} coupled problem for relatively high concentrations, i.e., not only the fibres are transported and deformed by the flow, but the flow in turns gets altered by their backreaction due to the no-slip condition at the surface of the immersed objects~\citep{balachandar2010turbulent,maxey2017simulation,brandt2021particle}.  Modelling such dynamical feedback by means of suitable techniques, e.g., immersed boundary methods, poses significant demands which are becoming more affordable with the availability of more powerful computational resources. Besides, along with ensuring the mutual coupling between the two phases, such fully-resolved approaches are expected to give more accurate results compared with traditional models used for describing the dynamics of the dispersed phase, whose foundation rely on one-way coupling and often linear flow assumptions~\citep{jeffery1922motion}.  This is especially the case when adopting the latter beyond its limit of applicability for anisotropic particles of finite-size, i.e., larger than the Kolmogorov lengthscale. 

The modulation of turbulence in non-dilute conditions has been the subject of studies regarding several classes of particulate and multiphase flows, e.g., spherical and isotropic particles~\citep{lucci2010modulation,gualtieri2013clustering,capecelatro2018transition,ardekani2019turbulence,ardekani2019turbulent,sozza2020drag}, anisotropic particles and fibres~\citep{andersson2012torque,ardekani2017drag,olivieri2020dispersed,olivieri2021universal}, as well as droplets or bubbles~\citep{dodd2016interaction,freund2019wavelet,rosti2019droplets,cannon2021effect}. Overall, when the concentration is large enough, this modulation effect typically causes a substantial departure from the classical phenomenology observed for a purely Newtonian fluid (e.g., the presence of a clear energy cascade for sufficiently large Reynolds number). Specifically, some common features are generally observed looking at the turbulent energy spectra despite the remarkable differences between these multiphase flows~\citep{gualtieri2013clustering,dodd2016interaction,rosti2019droplets, olivieri2021universal}: with respect to the reference single-phase case, it is typically found \textit{(i)} a decrease of the energy content at the lowest wavenumbers (i.e., the largest, energy-containing scales), along with \textit{(ii)} a relative enhancement of energy at higher wavenumbers (i.e., smaller scales).  Nevertheless, both the qualitative and quantitative comprehension on the mechanisms underlying such complex energy redistribution process is far from being exhaustive. Furthermore,  these effects are not only directly associated with the resulting flow dynamics, but are in turns potentially influential on the behaviour of the dispersed objects. 

\begin{figure}
\centerline{\includegraphics[width=0.62\textwidth]{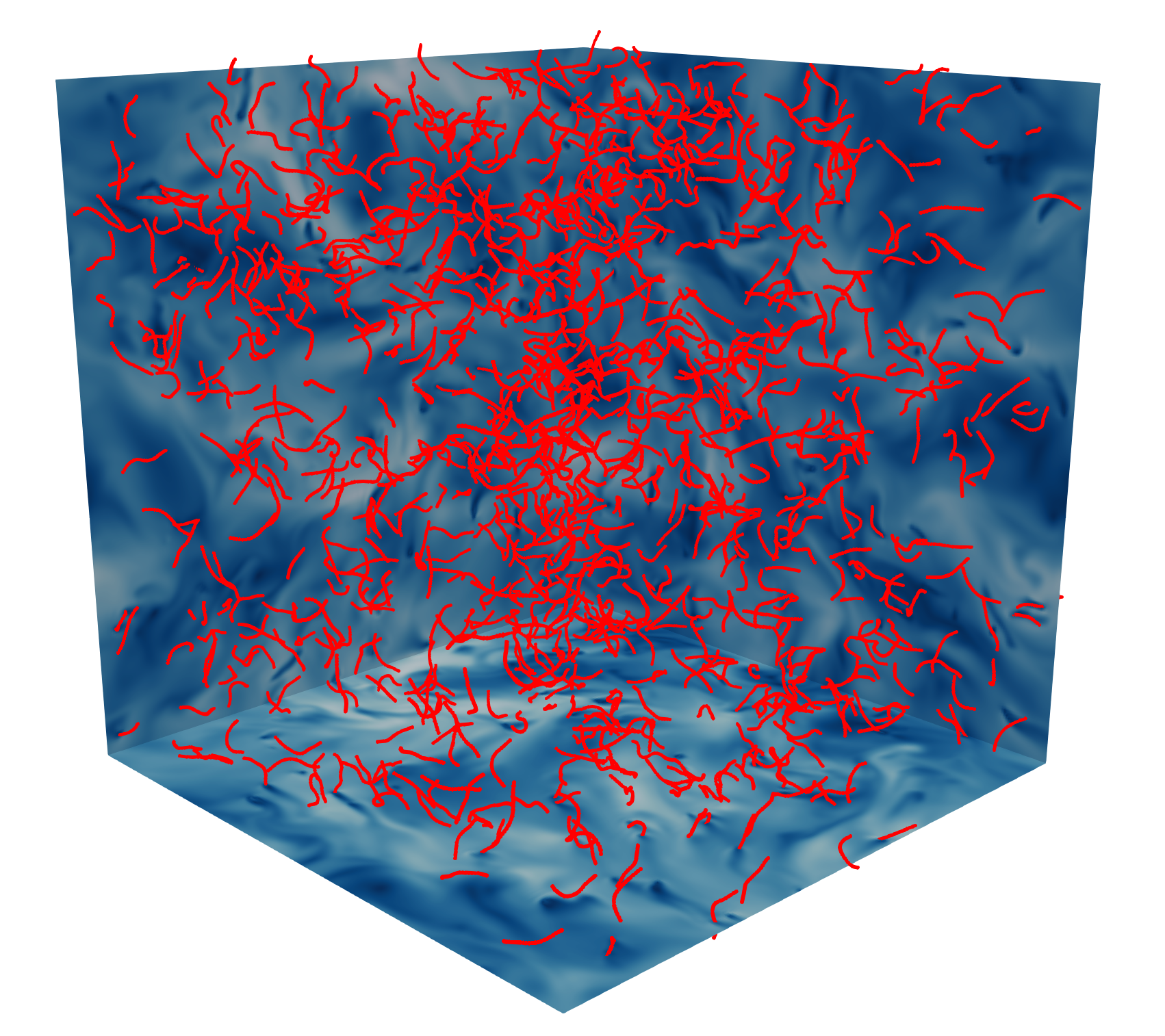}}
\caption{Snapshot from a direct numerical simulation of homogeneous isotropic turbulence where finite-size, flexible fibres (in red) are dispersed.  In particular, we show a high concentration case with very flexible \added{denser-than-the-fluid} fibres of intermediate length comparable to the inertial range of scales (i.e., \added{$\Delta \widetilde{\rho} = 1$}, $c=0.5$, $\gamma=10^{-8}$ and $N=10^3$). The domain backplanes are coloured by the fluid velocity magnitude.}
\label{fig:Qcr-and-fibers}
\end{figure} 

In this work, we investigate numerically the behaviour of a suspension of finite-size, inertial fibres, either rigid or flexible, dispersed in a homogeneous and isotropic turbulent flow in both dilute and non-dilute conditions. An illustrative example of the system under consideration is shown in \cref{fig:Qcr-and-fibers}. A vast parametric study is conducted over the main mechanical and geometrical properties of the fibres, i.e., linear density, length and bending stiffness, as well as their number\added{, while neglecting the presence of gravity}. In order to accurately describe the dynamics of fibres, whose length is well beyond the Kolmogorov lengthscale, and to take into account the backreaction of the dispersed phase to the carrier flow, a \added{\textit{four-way} coupled}, direct numerical simulation (DNS) approach is employed, based on the combination of the finite difference and immersed boundary methods\added{, and complemented by a fiber-to-fiber interaction model}. The goal of the work is twofold:
 \textit{(i)} to properly characterise the backreaction effect exerted by fibres to the turbulent flow, highlighting the role of the mass fraction as the representative control parameter, and providing a scale-by-scale description of the energy transfer across the scales of the fluid motion;
 \textit{(ii)} to describe various aspects of the dynamics of the dispersed fibres in the turbulent flow, including the existing flapping states, as well as the preferential concentration (i.e., tendency to clustering) and orientation (i.e., alignment with specific quantities of the flow).  \added{Note that this investigation partially recovers the results presented by~\citet{olivieri2021universal}, with the goal of substantially advancing such findings and offer a systematic and fully exhaustive analysis.}
 
The rest of the paper is structured as follows: in \cref{sec:methods}, we introduce the computational methodology used in our investigation along with providing information on the performed parametric study; in \cref{sec:results}, we present and discuss the results to address the two main objectives; finally, concluding remarks are given in \cref{sec:conclusions}.

\section{Methods}
\label{sec:methods}
This section describes the methodology adopted for the present study\added{, which is essentially the same adopted by~\citet{olivieri2021universal}}. First, in \cref{sec:problem}, we introduce the governing equations and main parameters of the problem. Then, in \cref{sec:numerical}, we describe how the former are solved numerically. Finally, in \cref{sec:parametric}, we provide details on the performed parametric DNS study.

\subsection{Problem formulation}
\label{sec:problem}
We consider an ensemble of $N$ fibres freely moving in homogeneous and isotropic turbulence (\cref{fig:Qcr-and-fibers}). The fibre's length $c$ is varied with respect to the characteristic length scales of the turbulent flow, ranging from being of the order of the dissipative (or Kolmogorov) scale to that of the integral scale. Fibres are modeled as one-dimensional slender objects, i.e., their length is assumed much larger than the diameter $d$, or equivalently we consider an aspect ratio $c/d \gg 1$. Moreover, we assume that the fibres are inextensible and focus solely on the elastic contribution due to the bending stiffness $\gamma$ (which for a homogeneous fibre is given by the product of the elastic modulus and the second moment of the area). Finally, $\Delta \widetilde{\rho} = \widetilde{\rho}_\mathrm{s} - \widetilde{\rho}_\mathrm{f}$ is the difference in the linear density\footnote{\added{Linear and volumetric densities are denoted with and without the tilde, respectively.}} of the fibre and the fluid, i.e., the parameter controlling the inertia of the dispersed objects. In particular, in the limit $\Delta \widetilde{\rho} \to 0$ we recover the \added{iso-dense} case.
\added{Here, we assume zero-gravity conditions in order to decrease the number of independent parameters. In particular, we note that the choice is justified when focusing on the limit of arbitrarily large Froude number.}

The dynamics of each fibre evolves according to the Euler-Bernoulli beam equation for an elastic filament coupled with the inextensibility constraint, which read
\begin{equation}
\Delta \widetilde{\rho} \,  \ddot{\Xb}= \partial_s \left( T \partial_s \Xb \right) - \gamma \partial^4_s \Xb - \textbf{F}_\mathrm{fs} + \textbf{F}_\mathrm{col},
\label{eq:EB1}
\end{equation}
\begin{equation}
\partial_s \Xb \cdot \partial_s \Xb = 1,
\label{eq:EB2}
\end{equation}
where $\textbf{X} \left( s, t \right)$ is the position of a generic material point belonging to the fibre, $T \left( s,t \right)$ is the tension enforcing the inextensibility constraint in the filament equation,  $\textbf{F}_\mathrm{fs} \left( s, t \right)$ is the fluid-solid interaction forcing and $\Fb_\mathrm{col} \left( s, t \right)$ is a fibre-to-fibre collision forcing term (both forcings will be discussed in the following); all quantities are a function of the curvilinear coordinate $s$ and time $t$. Freely-moving boundary conditions are imposed at both fibre's ends: $ \partial_{ss} \Xb |_{s=0,c} = 0$, $\partial_{sss} \Xb |_{s=0,c} = 0$ and $T |_{s=0,c} = 0$. From a normal mode analysis in the case of such configuration (i.e. free-free or unsupported beam) we can readily obtain the natural frequency $f_\mathrm{nat} = \alpha \, \sqrt{\gamma / (\Delta \widetilde{\rho} c^4)}$, where $\alpha \approx 22.4/\pi$. Dealing with finite-size objects, we neglect the Brownian contribution to the fibre dynamics, which is considered to be small compared to the hydrodynamic forcing. 

Fibres are released within a cubic domain of size $L=2\pi$ having periodic boundary conditions in all directions and where a homogeneous and isotropic turbulent flow is generated using the stochastic spectral forcing by~\citet{eswaran1988forcing}, randomly injecting energy at the largest scales, i.e., within a low-wavenumber shell with $1\le k \le2$ (with $k$ denoting the wavenumber). The carrier flow is governed by the incompressible Navier-Stokes equations for a Newtonian fluid, which read
\begin{equation}
\partial_t \ub + \ub \cdot \partialb \ub = - \partialb p / {\rho_\mathrm{f}} + \nu \partial^2 \ub + \fb_\mathrm{tur} +  \fb_\mathrm{fs},
\label{eq:NS1}
\end{equation}
\begin{equation}
\partialb \cdot \ub = 0,
\label{eq:NS2}
\end{equation}
where $\ub \left( \xb,t \right)$ and $p\left( \xb,t \right)$ are the velocity and pressure fields, respectively, $\rho_\mathrm{f}$ is the volumetric fluid density and $\nu$ is the kinematic viscosity; $\fb_\mathrm{tur} \left( \xb,t \right)$ is the external forcing used to sustain the turbulent flow~\citep{eswaran1988forcing} and $\fb_\mathrm{fs} \left( \xb,t \right)$ is an additional forcing mimicking the presence of the solid phase (as described in the following); lastly, $\xb$ denotes the position vector in the Eulerian framework.

The flow and fibre dynamics are coupled by the no-slip condition $\dot{\Xb} = \ub \left( \Xb\left( s, t \right),t \right)$. 

\subsection{Numerical technique}
\label{sec:numerical}
To solve numerically the fully-coupled FSI problem, we employ the IBM procedure originally proposed by~\citet{huang_shin_sung_2007a} and subsequently modified by~\citet{banaei2019numerical} and~\citet{olivieri2020dispersed,olivieri2020turbulence,olivieri2021universal}. In particular, the mutual interaction between the fluid and solid phase is enforced indirectly by means of the singular force distributions $\Fb_\mathrm{fs}$ and $\fb_\mathrm{fs}$ acting on the fibre and flow, respectively~\citep{peskin2002,huang_shin_sung_2007a}. The fluid velocity at the position of the fibre Lagrangian point, $\textbf{U} = \ub \left( \textbf{X} \left( s,t \right),t \right)$, is obtained by interpolating the fluid velocity at the Eulerian nodes surrounding the Lagrangian point:
\begin{equation}
\textbf{U} \left( \textbf{X} \left( s,t \right),t \right) = \int \textbf{u} \left( \textbf{x},t \right) \delta \left( \textbf{x} - \textbf{X}\left( s,t \right) \right) \mathrm{d}^3 \textbf{x},
\end{equation}
where $\delta$ is the Dirac delta function. The interpolated velocity $\Ub$ is used to compute the fluid-solid interaction forcing needed to enforce the no-slip condition
\begin{equation}
\Fb_\mathrm{fs}(s,t) = \beta \, (\Ub - \dot{\Xb}),
\end{equation}
where $\beta$ is a large negative constant~\citep{huang_shin_sung_2007a}. Finally, $\Fb$ is spread to the surrounding fluid flow as 
\begin{equation}
\fb_\mathrm{fs} \left( \textbf{x},t \right) = \frac{1}{{\rho}_\mathrm{f}} \int_s \Fb_\mathrm{fs} \left( s,t \right) \delta \left( \textbf{x} - \textbf{X}\left( s,t \right) \right) \mathrm{d} s.
\end{equation}
Both the interpolation and spreading operations feature the Dirac operator. Numerically, this is transposed into the use of the regularized $\delta$ function proposed by~\citet{roma1999}.
Consequently, in the numerical framework the fibre's diameter is also approximated to a finite value comparable to the grid spacing \added{(note that the present method is not able to properly resolve the boundary layer around the individual fibre diameter)}; in the following we assume $d = 2 \, \Delta x$ as the effective diameter.

Each fibre is evenly discretized using $N_\mathrm{L}$ Lagrangian points such that the spatial resolution $\Delta s = c / (N_\mathrm{L} -1)$ is approximately equal to the Eulerian grid spacing $\Delta x$; consequently, the discretized curvilinear coordinate can be evaluated as $s_l = l \, \Delta s$, with $l = 1, \dots, N_\mathrm{L}$. The numerical solution of~\cref{eq:EB1,eq:EB2} follows the scheme detailed in~\citet{huang_shin_sung_2007a} with the difference that the bending term is treated implicitly to allow for a larger timestep~\citep{banaei2019numerical,olivieri2021universal}. Furthermore, a fibre-to-fibre collision model is implemented to prevent that the distance $\mathbf{d}$ between Lagrangian points belonging to different fibres goes below the grid spacing $\Delta x$ and that fibres eventually cross each other. Specifically, we employ the minimal collision model proposed by~\citet{snook2012vorticity}, where a constant forcing $\Fb_\mathrm{col} = F_0 \, \mathbf{\hat{d}}$ is applied when $|\mathbf{d}| \le \Delta_\mathrm{col}$. Here, $F_0$ is a free parameter and $\Delta_\mathrm{col} = \mathcal{O}(\Delta x)$ is the critical distance below which the forcing is imposed. After several tests on the sensitivity of the results with respect to these parameters (both in simple laminar flow conditions with few fibres and in the turbulent case with a non-dilute suspension), we chose $F_0 = 1.0$ and $\Delta_\mathrm{col} = 3 \Delta x$. \added{As a result, the influence of collisions turns out to be very limited, as shown in \cref{app:collisions}: fiber-to-fibre interactions are extremely rare in the majority of the analyzed configurations and are always found to have a negligible effect on the numerical results, especially on the turbulence modulation and fluid-solid coupling mechanisms which are the main focus of this work.}

Concerning the fluid flow, \cref{eq:NS1,eq:NS2} are solved using the fractional step method on a staggered grid~\citep{kim2007penalty}. The Poisson equation enforcing the incompressibility constraint is solved using a fast and efficient approach based on the Fast Fourier Transform (FFT). The numerical solution is based on the (second-order) central finite-difference method for the spatial discretization and the (third-order) Runge-Kutta scheme for the temporal discretization. 

The described computational procedure is implemented in the in-house solver \textit{Fujin} (\href{https://groups.oist.jp/cffu/code}{\texttt{https://groups.oist.jp/cffu/code}}). The code is parallelized using the MPI protocol and the \textit{2decomp} library for domain decomposition (\href{http://www.2decomp.org}{\texttt{http://www.2decomp.org}}). It has been extensively tested and employed in a variety of fluid dynamical problems~\citep{rosti2019droplets,rosti2020increase,rosti2020fluid,rosti2021turbulence}, including in particular suspensions of fibres in both laminar and turbulent flows~\citep{rosti2018flexible,rosti2019flowing,cavaiola2019assembly,olivieri2020turbulence,olivieri2020dispersed,olivieri2021universal,brizzolara2021fibre}. To further enrich the existing \added{framework}, in~\cref{sec:validation} \added{we show a comparison of} our results on the self-sustained flapping oscillation of a two-dimensional, hinged filament in a laminar flow with those originally reported by~\citet{huang_shin_sung_2007a}.

\subsection{Parametric study}
\label{sec:parametric}
The objective of this work is to perform an extensive parametric study over the characteristic properties of the fibre suspension. Therefore, we have performed DNS varying the fibre's linear density difference $\Delta \widetilde{\rho}$, length $c$ and bending stiffness $\gamma$, along with the number of fibres $N$. Overall, this also leads to consider different concentrations of the suspension, i.e., from very dilute to non-dilute suspensions. Specifically, in a first baseline investigation we considered two different linear densities $\Delta \widetilde{\rho}$, corresponding to essentially \added{iso-dense} and \added{denser-than-the-fluid} fibres, three different values of the fibre's length $c$, which are comparable to the dissipative, inertial and integral characteristic lengthscales of the turbulent flow,  three different bending stiffnesses $\gamma$ so that we range from highly flexible to essentially rigid fibres. Finally, we varied the numbers of fibres $N$ to adjust the concentration of the suspension. \Cref{tab:caselist_nb,tab:caselist_in,tab:caselist_in2} in \cref{sec:list} report the parametric combinations along with the consequent relevant metrics for the concentration, e.g., number density or mass fraction, and the corresponding symbol used in the rest of the paper to refer to each case.

A reference configuration is chosen in the single-phase case (i.e., for $N=0$), where the turbulent flow is characterized by a micro-scale Reynolds number $\mathrm{Re}_\lambda \equiv u_\mathrm{rms} \lambda / \nu \approx 120$, where $u_\mathrm {rms}$ is the root-mean-square velocity and $\lambda$ is the Taylor micro-scale. To simulate such configuration, the fluid domain is discretized into a uniform Eulerian grid using $256^3$ cells, having verified that the ratio between the Kolmogorov dissipative lengthscale and the grid spacing is $\eta/\Delta x = \mathcal{O}(1)$. Moreover, we verified the convergence of the numerical results by doubling the spatial resolution in one representative case (not shown), observing negligible differences with respect to the adopted grid setting. After reaching the statistically stationary state in the single-phase configuration, the obtained flowfield is used as the initial condition for the multiphase flow simulations, i.e., fibres are added to the fully-developed turbulent carrier flow. The simulation is then evolved to reach the new stationary regime, the achievement of the latter being qualitatively assessed from the time histories of main statistical quantities for both phases (e.g., average kinetic energy). Hence, we disregard the transient and accumulate the statistics over a minimum of $5$ integral timescales for all cases.  The statistical convergence of our data concerning both the flow and fibre dynamics was verified by widely varying the number of samples and checking their sensitivity on the results, ensuring that all the reported differences are substantially larger than the statistical uncertainty.

\begin{figure}
\centering
\includegraphics{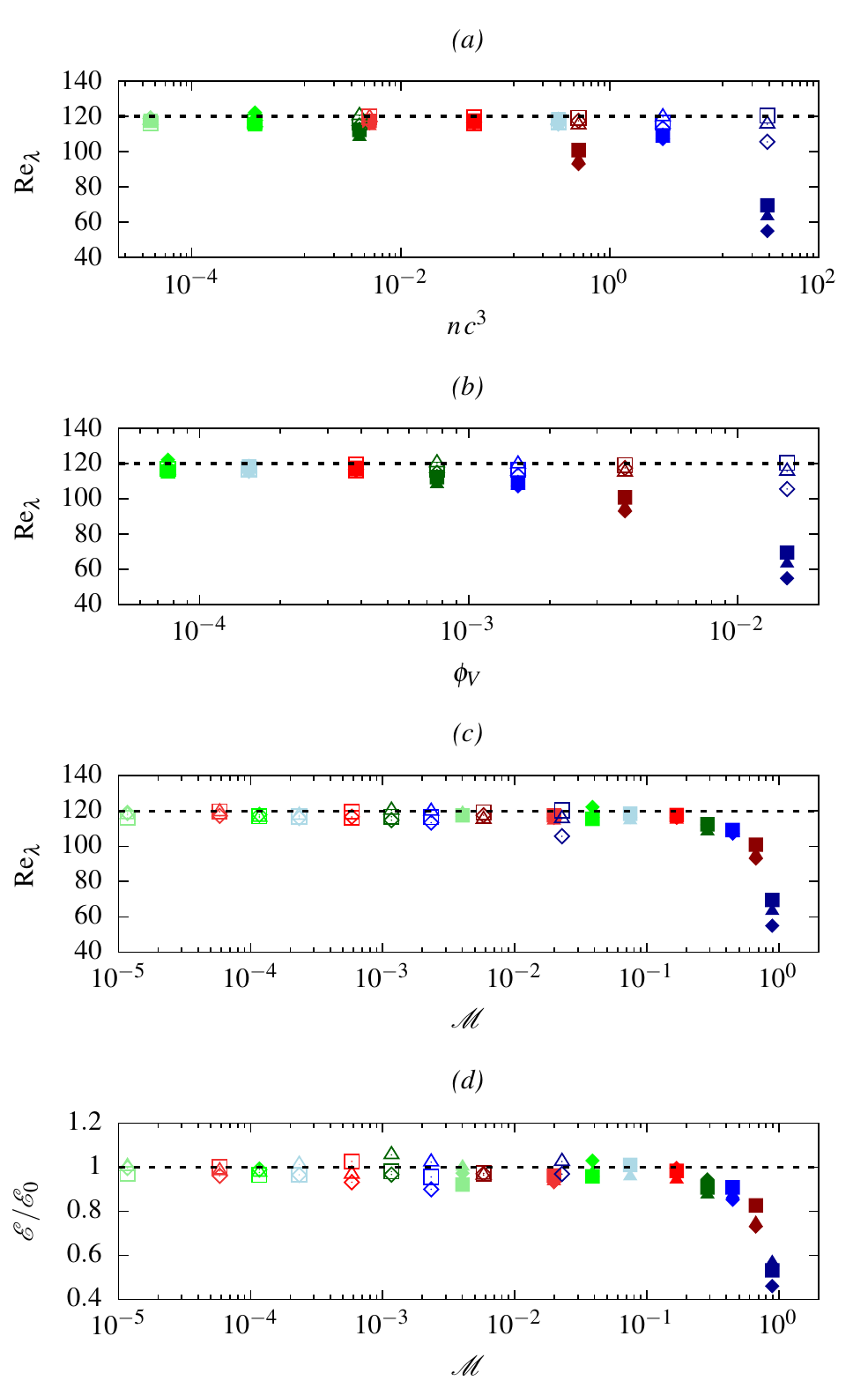}
\caption{\added{Micro-scale Reynolds number as a function of the \textit{(a)} (nondimensional) number density, \textit{(b)} volume fraction and \textit{(c)} mass fraction (each of these indicator quantities are defined in the text; the dashed line in each plot indicates the reference value obtained in the single-phase case), and \textit{(d)} turbulent kinetic energy (normalized with the value of the single-phase case) as a function of the mass fraction.}  Colors and symbols are used consistently with the indication of \cref{tab:caselist_nb,tab:caselist_in}:
empty and filled markers denote the cases for \added{iso-dense} ($\Delta \widetilde{\rho} = 10^{-3}$) and \added{denser-than-the-fluid} ($\Delta \widetilde{\rho} = 10^0$) fibres, respectively.
The colour denotes the fibre's length (green: $c=0.1$, i.e. short; red: $c=0.5$, i.e. intermediate; blue: $c=2.0$, i.e. long).
The symbol indicates the bending stiffness (squares: $\gamma=10^{-8}$, triangles: $\gamma=10^{-4}$, rotated squares: $\gamma=10^{0}$).
Finally, the brightness decreases with the number of fibres (light: $N=10^1$, medium: $N=10^2$, dark: $N=10^3$).}
\label{fig:ReL}
\end{figure}

\section{Results}
\label{sec:results}
This section presents the results of our study. First, in \cref{sec:flow-dyn}, we focus on the characterisation of the modulation of the turbulent flow due to the dispersed phase. Then, in \cref{sec:fiber-dyn}, we analyse in more detail several features of the fibre motion and transport.

\subsection{Flow dynamics}
\label{sec:flow-dyn}

\subsubsection{Macroscopic behaviour}

We begin our analysis by looking at the bulk properties of the turbulent flow\deleted{, and specifically the micro-scale Reynolds number $\mathrm{Re}_\lambda$,} when varying the parameters of the suspension. \added{To start,} \cref{fig:ReL}\added{\textit{a,b,c}} shows the dependence of \added{the micro-scale Reynolds number} $\mathrm{Re}_\lambda$ as a function of several quantities that can be used as an indicator to characterise the concentration of the suspension: (\textit{i}) the (nondimensionalized) number density $n \, c^3$, (\textit{ii}) the volume fraction $\phi_V$, and (\textit{iii}) the mass fraction $\mathcal{M}$. These three quantities are chosen since they are often used in the framework of fibre suspensions, or more generally particle-laden flows, to estimate (a priori) the importance of the backreaction of the dispersed phase on the carrier flow~\citep{butler2018microstructural,brandt2021particle}. From \cref{fig:ReL}, it can be noted that for several cases the Reynolds number turns out to decrease with respect to the single-phase case (the value of the latter indicated by the horizontal dashed line in the figure); this appears as a robust effect that will be characterised in detail in the rest of the section. However, the ways in which the data appear when plotted as a function of the three different parameters are radically different.

The (dimensional) number density, defined as $n = N / V_\mathrm{f}$ (where $V_\mathrm{f} = L^3$ is the volume of the domain), is a quantity widely used to investigate fibre suspensions, especially in the framework of rheological studies in low-Reynolds-number flows. Such quantity is typically made nondimensional with the fibre length $c$, so that $n c^3$ represents a measure of the relative spacing between fibres and consequently of the degree of their mutual interaction. More specifically, the suspension is typically considered to be in the dilute regime if $n c^3 \ll 1$, whereas it is classified as semi-dilute for $ 1/c^{3} \leq n  \ll 1/ (d \, c^2)$. If $n  \gg 1/ (d \, c^2)$, i.e., the spacing between the fibres is below their diameter, we finally have the concentrated regime, where the fibre-to-fibre interactions are expected to be dominant in controlling the dynamics of the dispersed objects~\citep{butler2018microstructural}. We note that all the configurations considered in the present study fall either in the dilute or in the semi-dilute regime. As shown in~\cref{fig:ReL}\textit{a}, however, the number density does not appear to be a good indicator for describing the entity of the backreaction on the turbulent flow, without any systematic trend observed when plotting the numerical results with respect to such parameter. As it will be clearer in the following, the reason for this is the incorrect scaling with the fibre length.

Let us now consider the same data but as a function of the volume fraction $\phi_V$, as reported in~\cref{fig:ReL}\textit{b}. Here, $\phi_V = V_\mathrm{s} / V_\mathrm{f}$ is defined as the ratio between the volume of the dispersed phase $V_\mathrm{s} = N \, c \, \pi d^2 / 4$ and that of the fluid domain $V_\mathrm{f} = L^3$. Using this quantity, the results are outlined with a much more defined trend (compared with those for the number density), with $\mathrm{Re}_\lambda$ decreasing as $\phi_V$ is increased. However, it can be noted that such indicator misses to take into account the inertia of the fibres, here quantified by the linear density difference $\Delta \widetilde{\rho}$. Indeed, the latter clearly appears as an additional key (and free) parameter, with a dramatic difference between the results for the \added{iso-dense} (empty symbols) and the \added{denser-than-the-fluid} (filled symbols) fibre cases. 

In light of the evidence above, it is therefore natural to consider, as another way to parametrize the backreaction effect, the mass fraction $\mathcal{M} = m_\mathrm{s} / (m_\mathrm{s} +m_\mathrm{f})$, where $m_\mathrm{s} = \rho_\mathrm{s} V_\mathrm{s}$ is the (total) mass of the solid dispersed phase and $m_\mathrm{f} = \rho_\mathrm{f} \, V_\mathrm{f}$ is the (total) mass of fluid in the domain. As a result, from \cref{fig:ReL}\textit{c} it can be now appreciated how the mass fraction turns out to be the most representative parameter for describing the variation of the backreaction and consequent turbulence modulation at a macroscopic level. We highlight that, although this fact is well known for the case of point-like inertial particles~\citep{brandt2021particle}, similar evidence is not reported for suspensions of finite-size fibres, such as those investigated in the present work. Focusing on \cref{fig:ReL}\textit{c}, it can be observed that the Reynolds number starts to vary appreciably from the single phase value when the mass fraction becomes larger than around $10\%$, thus systematically decreasing with $\mathcal{M}$ and eventually reaching, for the highest mass fraction that was tested ($\mathcal{M}\approx89\%$), a value that is roughly half of that obtained in the single-phase case ($\mathrm{Re}_\lambda \approx 60$).

\added{For a more complete characterisation of the turbulence modulation, in \cref{fig:ReL}\textit{d} we also show how the turbulent kinetic energy decreases with the mass fraction, following a trend that is qualitatively similar to what observed for the micro-scale Reynolds number.}

From \cref{fig:ReL}, a further comment can be made on a secondary yet systematic effect associated with the variation of the fibre's bending stiffness, which is especially evident for the longest fibres at the highest concentration (dark blue symbols). Remarkably, the Reynolds number is systematically found to decrease when the bending stiffness $\gamma$ is increased (the fibre is more rigid). This feature is particularly evident for the cases with the longest fibres and can be intuitively associated with the degree of compliance manifested by the elastic objects under the action of the flow, which reflects in a stronger friction exerted on the flow.
Nevertheless, it can be also pointed out the huge variation (up to eight orders of magnitude) in the bending stiffness that was considered for the reported cases, in light of which we conclude that the effect associated with such parameter is overall largely subleading when compared to that caused by the inertia of the dispersed phase.

\subsubsection{Energy spectra} 
\begin{figure}
\centering
\includegraphics{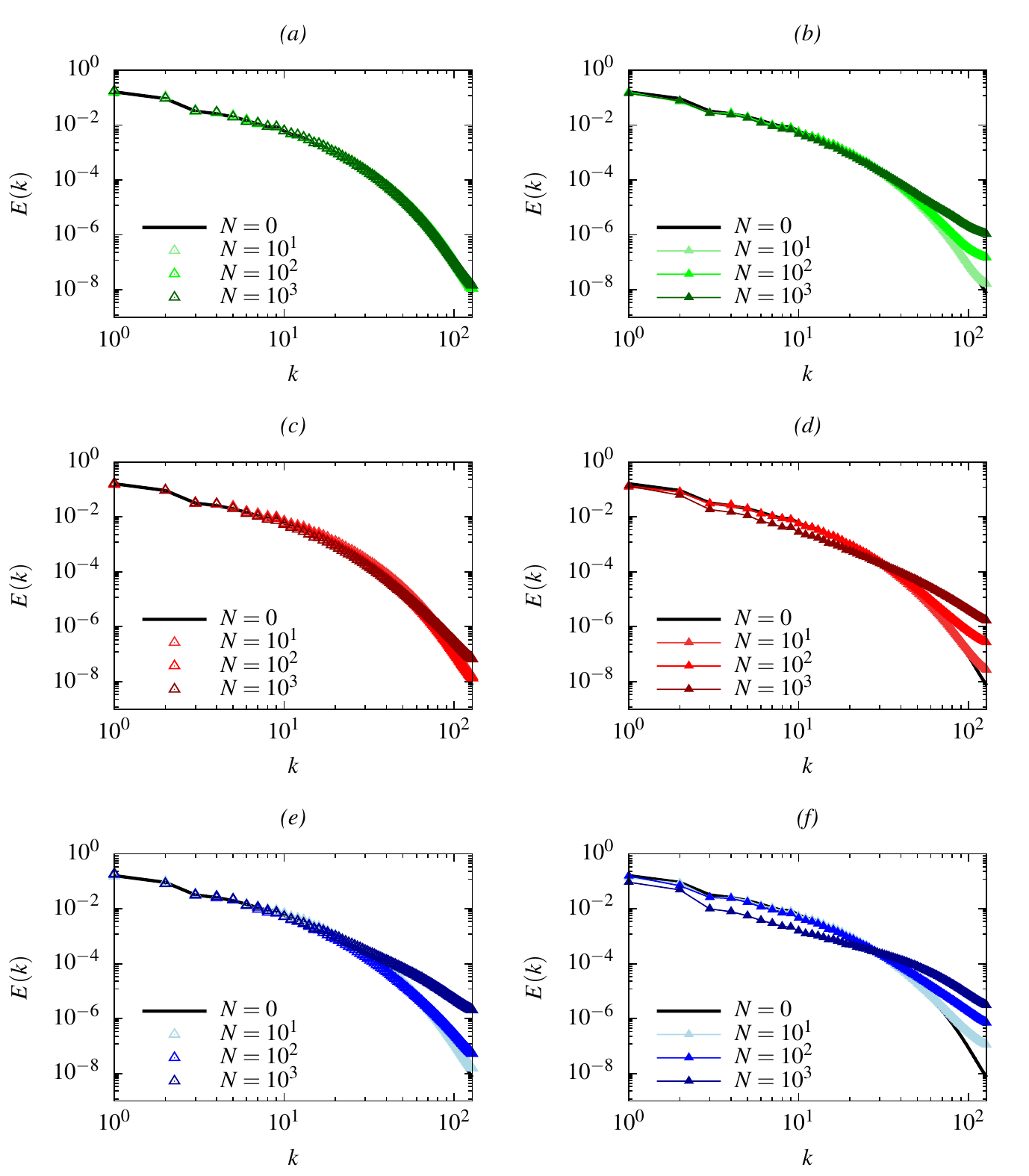}
\caption{Energy spectra of the modulated turbulent flow for different fibre suspension. Left \textit{(a,c,e)}: \added{iso-dense} fibres; right \textit{(b,d,f)}: \added{denser-than-the-fluid} fibres. Top \textit{(a,b)}: short fibres; middle \textit{(c,d)}: intermediate fibres; bottom \textit{(e,f)}: long fibres. The number of fibres, and therefore the mass fraction, increases from light to dark. Colors and symbols are used consistently with \cref{tab:caselist_nb,tab:caselist_in}. All cases are relative to the intermediate value of the bending stiffness. The black lines (without symbols) refer to the single-phase case.
}
\label{fig:spectra_inert-vs-neutr}
\end{figure}

To better understand the multiscale features of the backreaction of the dispersed phase on the carrier flow, in~\cref{fig:spectra_inert-vs-neutr} we show the energy spectra computed from our simulations for the two linear densities $\Delta \widetilde{\rho}$, i.e., \added{iso-dense} (left panels) vs \added{denser-than-the-fluid} fibres (right panels), and the three fibre lengths $c$ (corresponding to each row and increasing from top to bottom) investigated. In each plot,  we vary the number of fibres $N$ (increasing from light to dark color). Moreover, for the sake of comparison, the results from the single-phase case are also reported (black curve). Here, we retain the same intermediate value for the bending stiffness $\gamma$ since its influence on the turbulence modulation was previously shown to be subleading when compared to the combined variation of the other parameters determining the mass fraction of the suspension. 

\Cref{fig:spectra_inert-vs-neutr} shows that, as the concentration is increased, the energy spectrum is modified with the same qualitative behavior in all the cases. The latter can be essentially described as the combination of an energy content depletion at the largest scales (i.e., low wavenumbers), along with a (relative) enhancement at the smallest scales (i.e., high wavenumbers), with respect to the single-phase configuration. Moreover, the energy distribution in the intermediate range of scales is also appreciably modified, giving rise to a rather different phenomenology compared with that of classical turbulence. Clearly, the effect is more pronounced when increasing the fibre density and/or length, consistently with the increase in the mass fraction as outlined from the macroscopic characterisation of the bulk flow properties.

To highlight that qualitatively similar flow conditions can be obtained using significantly different combinations of the suspension parameters, we can compare, e.g., the case of short, \added{denser-than-the-fluid} fibres in \cref{fig:spectra_inert-vs-neutr}\textit{b} with that of long, \added{iso-dense} fibres in \cref{fig:spectra_inert-vs-neutr}\textit{e} (considering for both cases the highest number of fibres). Although some quantitative differences are present (with the former showing a slightly stronger modulation), when compared to the single-phase case the resemblance between the two configurations is evident. Another observation can be inferred by looking at the enhanced energy content at the smallest cases, a robust feature manifesting already at relatively low mass fraction, and which can have potential relevance for mixing processes. On the other hand, it should be noted that the overall kinetic energy of the flow monotonically decreases with the mass fraction, consistently with the large-scale energy depletion previously discussed. 

\begin{figure}
\centering
\includegraphics{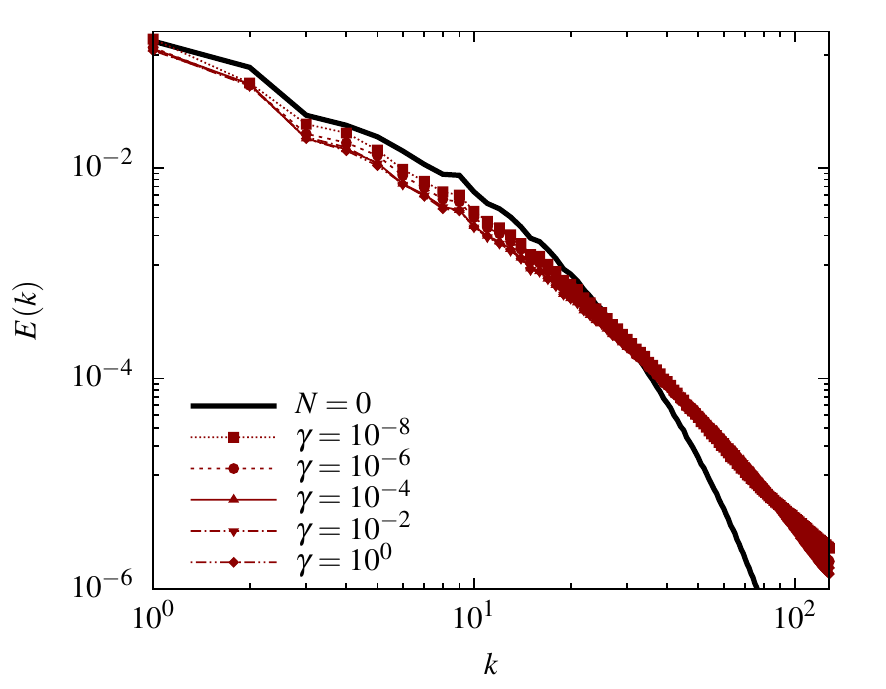}
\caption{
Effect of fibre's bending stiffness on the turbulent energy spectrum. Results are obtained for a suspension of $N=10^3$ \added{denser-than-the-fluid} fibres of intermediate length, varying the fibre's bending stiffness over eight orders of magnitude. The black curve refers to the single-phase case. Symbols are used accordingly to~\cref{tab:caselist_in2}.
}
\label{fig:spectra_gamma}
\end{figure}
 
Lastly, we explore the effect of the bending stiffness on the energy spectrum for one representative configuration (i.e.\added{, denser-than-the-fluid} fibres of intermediate length and maximum concentration\added{), shown in \cref{fig:spectra_gamma}.} 
On one hand, it can be noted that the energy content at the large scales, i.e. low wavenumbers, is systematically decreasing as the bending stiffness is increasing, in agreement with what observed in terms of the overall energy depletion (cf. \cref{fig:ReL}). On the other hand, in the high-wavenumber region the various curves are almost superimposed. 
\added{From the physical viewpoint, we can argue that increasing the flexibility (i.e., decreasing $\gamma$) leads to a more relaxed flow-structure coupling (i.e., fibres are more compliant to the action of the flow), reflected in a less pronounced backreaction if compared with the rigid case. Note that the saturation observed in \cref{fig:spectra_gamma} when increasing $\gamma$ corresponds indeed to the convergence towards the rigid configuration. In this regard, a more quantitative indication on the expected role of elasticity can be obtained in terms of the ratio between the structural and fluid characteristic timescales (later introduced in \cref{sec:fiber-dyn}), from which it can be deduced the extensive (nondimensional) range that has been considered (cf. \cref{fig:flapFreqRatio-vs-NatFreqRatio}).
Nonetheless, it has also to be pointed out that the variation with $\gamma$ appears overall very limited, confirming the secondary effect of flexibility in altering the carrier flow.
}

\begin{figure}
\centering
\includegraphics{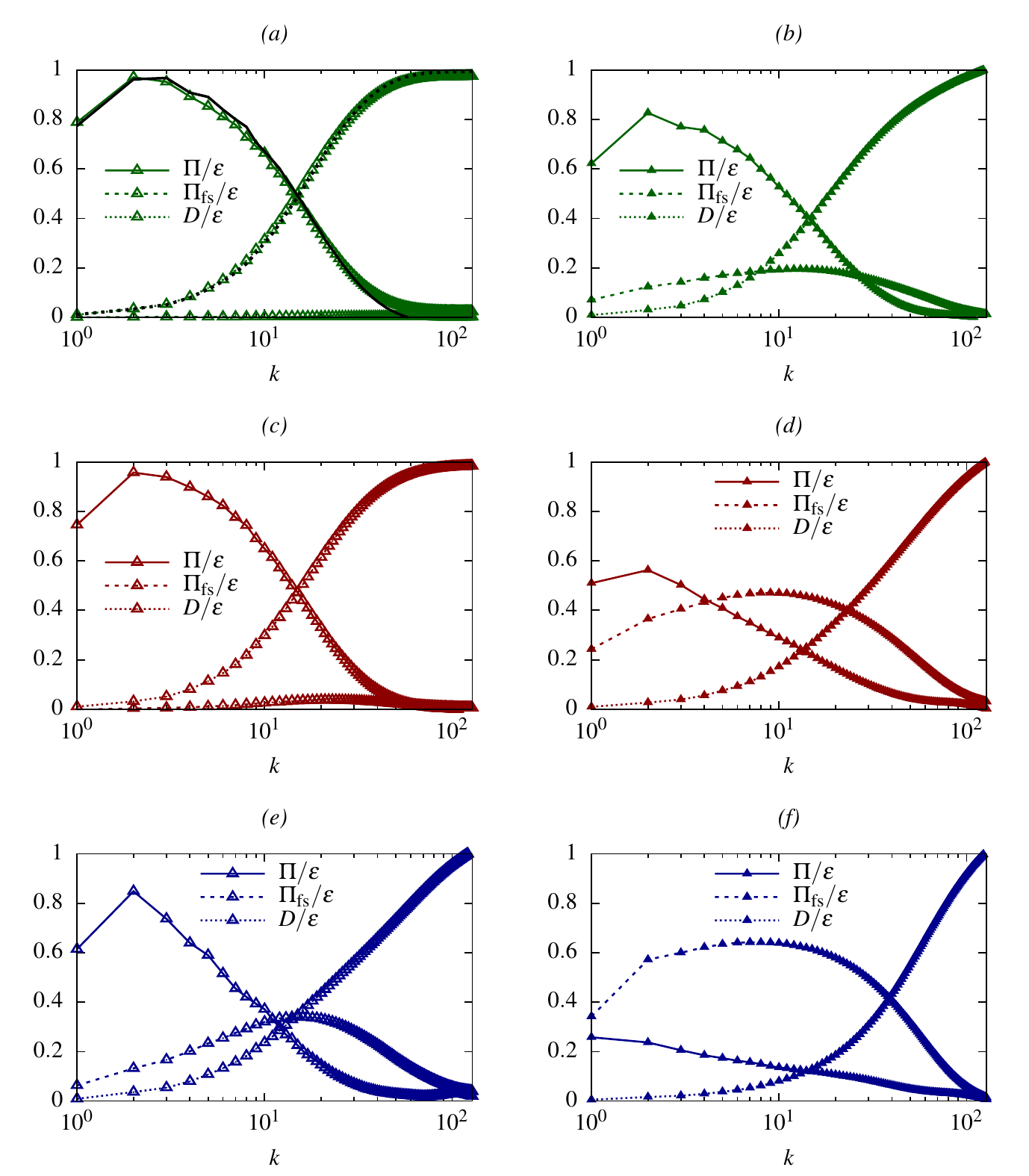}
\caption{Spectral power balance according to \cref{eq:spectral-power-balance}, showing the energy fluxes due to the nonlinear term (solid lines), fluid-solid coupling (dashed) and viscous dissipation (dotted), all normalized with the turbulent energy dissipation rate, for the cases with $N=10^3$ and intermediate bending stiffness.
Left \textit{(a,c,e)}: \added{iso-dense} fibres; right \textit{(b,d,f)}: \added{denser-than-the-fluid} fibres. Top \textit{(a,b)}: short fibres; middle \textit{(c,d)}: intermediate fibres; bottom \textit{(e,f)}: long fibres. In addition, in \textit{(a)} the nonlinear and dissipation terms of the single-phase case are also reported for comparison. 
Colors and symbols are used consistently with the indication of \cref{tab:caselist_nb,tab:caselist_in}. }
\label{fig:spectralFluxes_all0}
\end{figure}

\subsubsection{Scale-by-scale energy transfer} 
\label{sec:scale-by-scale}

In order to characterise the mechanisms underlying the outlined phenomenology, we can analyse the scale-by-scale energy transfer that for the present multiphase flow problem can be written as
\begin{equation}
 P(k) + \Pi(k) + \Pi_\mathrm{fs}(k) + D(k) = \epsilon,
\label{eq:spectral-power-balance}
\end{equation}
where the various terms appearing on the left hand side are the production rate $P$ (here associated with the external forcing used to sustain the flow and acting only at the largest scales, i.e., low wavenumbers), the energy flux $\Pi$ associated with the nonlinear term, the energy flux $\Pi_\mathrm{fs}$ associated with the fluid-solid coupling, and the viscous dissipation $D$. Note that the latter is defined such that the turbulent energy dissipation rate $\epsilon$ appears on the right hand side and the overall balance can be visualized more conveniently. For a derivation of \cref{eq:spectral-power-balance}, see \cref{app:spectral-balance}. In the single-phase case $\Pi_\mathrm{fs} \equiv 0$, and the dominant terms in the balance are the nonlinear term $\Pi$ and the viscous dissipation $D$ when approaching the lowest and highest wavenumbers, respectively~\citep{pope2000turbulent}.

In this analysis, we focus on the cases with $N=10^3$. However, the generality of the results is not restricted since the mass fraction, and consequently the entity of the backreaction, still varies considerably, thus ranging from configurations with negligible backreaction (i.e., essentially resembling the case without fibres) to configurations where the backreaction is strong and leads to a dramatic departure from the more classical scenario obtained for purely Newtonian fluid turbulence. Similarly, for the bending stiffness we select at first the same (intermediate) value as for the energy spectra reported in \cref{fig:spectra_inert-vs-neutr}, focusing on the effect of the mass fraction.

Let us therefore start from considering the spectral power balance for the cases of \added{iso-dense} fibres (with different fibre lengths), shown in \cref{fig:spectralFluxes_all0}\textit{a,c,e}. Here, the mass fraction is always relatively small; in agreement with the previously discussed effects on the Reynolds number and energy spectrum, we therefore expect an overall negligible, or at most quite limited, alteration in the balance with respect to the single-phase configuration. Indeed, starting from the short fibres (\cref{fig:spectralFluxes_all0}\textit{a}) it can be clearly observed that the results resemble those of the single-phase configuration (indicated by the black lines without symbols), with a predominance at low wavenumbers of the nonlinear term $\Pi$ and a negligible energy transfer associated with the fluid-structure coupling $\Pi_\mathrm{fs}$. This situation is accompanied by the predominance at high wavenumbers of the dissipation term $D$ which eventually saturates to $\epsilon$ for increasing $k$. The same applies also to fibres of intermediate length (\cref{fig:spectralFluxes_all0}\textit{c}), with only a very modest increase of $\Pi_\mathrm{fs}$.  On the other hand, when considering the case of long fibres (\cref{fig:spectralFluxes_all0}\textit{e}) the situation starts to be different: at low wavenumbers, the nonlinear term is weakened and does not entirely control anymore the energy transfer, with a non-negligible contribution of the fluid-solid coupling which is maximized at an intermediate lengthscale; as $k$ increases, the dissipation progressively becomes dominant, although the saturation is less evident due to the enhanced small-scale energy transfer. Remarkably, the mass fraction $\mathcal{M}$ in this case is only around $1\%$, therefore indicating that a substantially different and complex dynamics of the turbulent flow may arise already in relatively dilute conditions.

\begin{figure}
\centering
\includegraphics{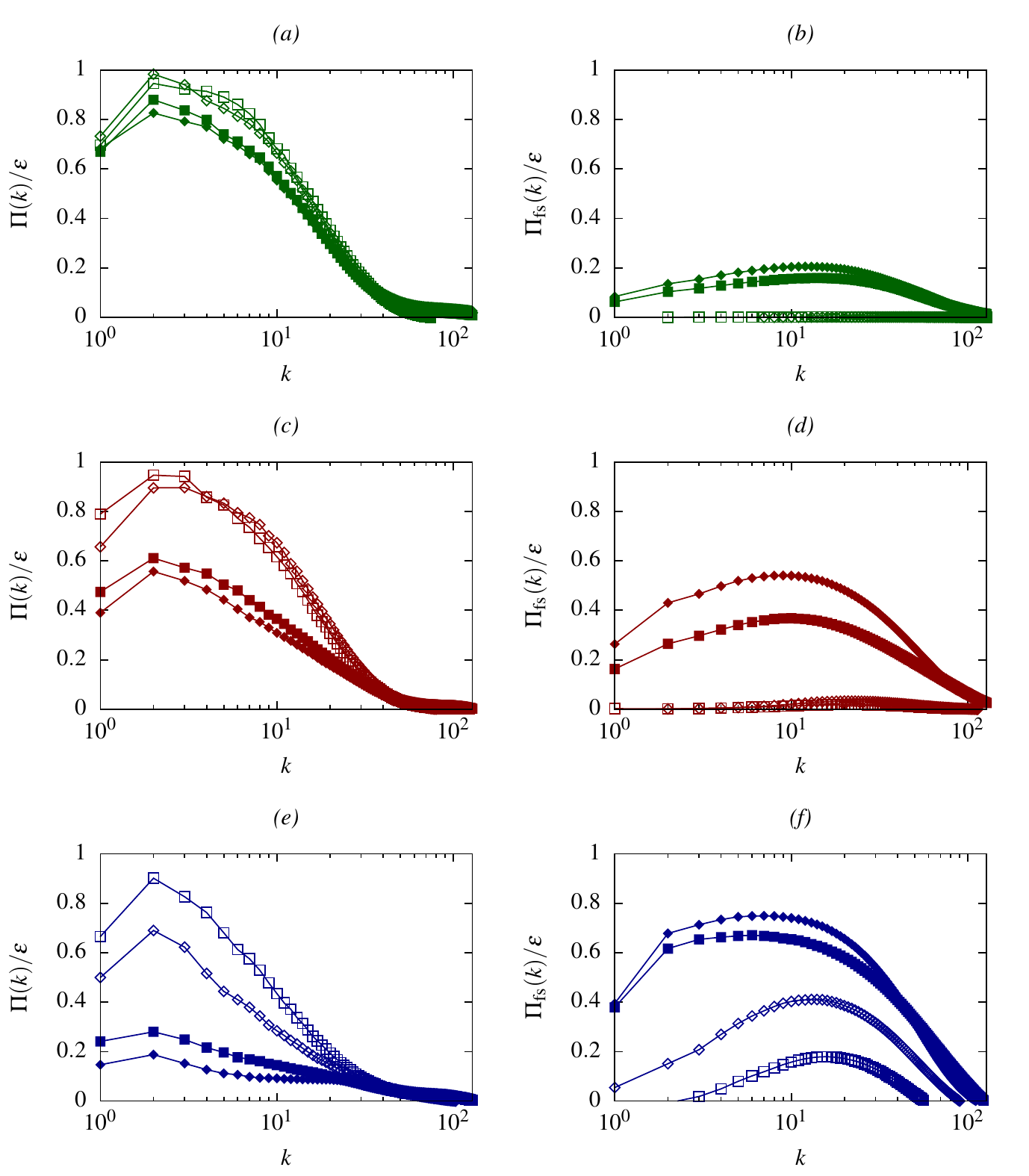}
\caption{Energy fluxes associated with the nonlinear term (left panels \textit{a,c,e}) and fluid-solid coupling term (right panels \textit{b,d,f}) for $N=10^3$ fibres with minimum vs maximum bending stiffness, along with different linear density and length. Colors and symbols are used consistently with the indication of \cref{tab:caselist_nb,tab:caselist_in}:
 Empty and filled markers refer to \added{iso-dense} and \added{denser-than-the-fluid} fibres, respectively.
 The symbol indicates the bending stiffness (squares: minimum, diamonds: maximum).
The colour denotes the fibre's length (green: short; red: intermediate; blue: long).
}
\label{fig:spectralFluxes_all}
\end{figure}

The alteration gets substantially more dramatic when considering the same three cases but for \added{denser-than-the-fluid} fibres, reported in \cref{fig:spectralFluxes_all0}\textit{b,d,f}. Already for short fibres (\cref{fig:spectralFluxes_all0}\textit{b}) a similar behavior can be noted, with a non-negligible fraction of the energy transfer that is associated to the fluid-solid coupling term. In fact, this case presented similarities already in terms of the energy spectrum with the one with \added{iso-dense} long fibres, as previously mentioned. When increasing the mass fraction and considering the intermediate fibre length (\cref{fig:spectralFluxes_all0}\textit{d}), the contribution of the nonlinear and fluid-solid coupling terms become comparable in magnitude. However, the former is dominant at lower wavenumbers, indicating that the energy from the largest scales is at first mainly transferred to the smaller scales by means of the classical hydrodynamic mechanism, although in a restricted subrange of scales compared with the single-phase case. Then, at sufficiently higher wavenumbers, such energy transfer is progressively replaced by another mechanism driven by the direct interaction between the carrier flow and the dispersed phase.We stress the fact that, as the mass fraction is increased, the nonlinear term decreases while the fluid-solid coupling is increased, while the viscous dissipation always eventually prevails at sufficiently large wavenumbers, i.e., small scales. Eventually, for the case of long \added{denser-than-the-fluid} fibres having the largest mass fraction (\cref{fig:spectralFluxes_all0}\textit{f}), the nonlinear term is essentially suppressed with the fluid-solid coupling largely controlling the overall energy transfer, up to the dissipative scales.

As an overall observation, for all cases the energy transfer is eventually compensated by the viscous term, since both the nonlinear and fluid-solid coupling term vanish when computed over the full range of wavenumbers. This aspect concerns a remarkable difference between the case of elastic fibres, as those considered in the present work, and that of polymer suspensions and viscoelastic turbulence, where the polymer stresses both transfer and dissipate energy~\citep{de2005homogeneous,valente2014effect}. In our model, no internal dissipation mechanism is considered for the dispersed phase and therefore the fluid-solid coupling purely transfers energy from larger to smaller scales, in a strict analogy with the nonlinear term, with the total energy flux given by the sum of the two contributions.

To conclude the analysis, there remains to discuss in more detail the influence of the fibre's flexibility, quantified by the bending stiffness $\gamma$, on the backreaction and turbulence modulation. \Cref{fig:spectralFluxes_all} reports the energy flux from the nonlinear (left panels) and fluid-solid coupling (right panels) contributions for the cases with maximum and minimum bending stiffness (along with different density and length) in order to highlight the overall range of variation associated with this parameter. The plots confirm once more that the variation is always quite modest, if compared with the one given by the other parameters (i.e., more universally expressed in terms of the mass fraction). Nevertheless, a systematic trend in all cases can be detected: as the bending stiffness is increased, the magnitude of the nonlinear term is found to decrease whereas the fluid-solid coupling contribution becomes more relevant. This effect, which could be intuitively expected, can be associated with the fact that rigid fibres are less compliant and do not easily adapt to the stresses applied by the turbulent flow in the same way as very flexible and more compliant fibres do. Remarkably, this mechanism is intrinsically different from the one governed by the inertia of the dispersed phase, and its potential relevance could be better outlined in very dense suspensions of \added{iso-dense} fibres.

\subsection{Fibre dynamics}
\label{sec:fiber-dyn}
\begin{figure}
\centering
\includegraphics{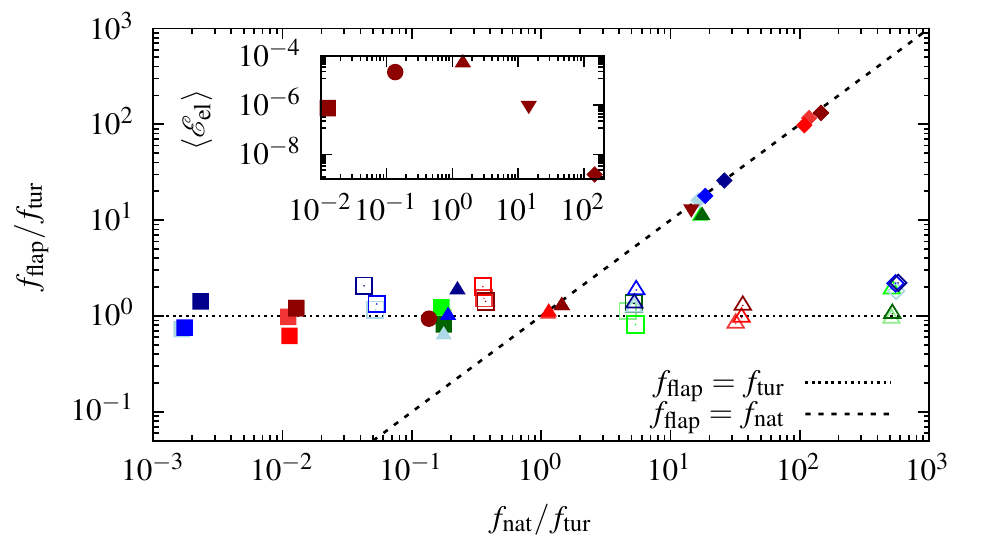}
\caption{Fibre's dominant flapping frequency as a function of the (free-response) natural frequency, both normalized with the effective eddy turnover frequency (evaluated at the fibre lengthscale). Dotted and dashed lines denote the expected scalings in the two flapping states. The inset reports the average elastic energy of the fibres for $N=10^3$ \added{denser-than-the-fluid} fibres of intermediate length and different bending stiffness. Colors and symbols are used consistently with the indication of \cref{tab:caselist_nb,tab:caselist_in,tab:caselist_in2}.}
\label{fig:flapFreqRatio-vs-NatFreqRatio}
\end{figure}

\subsubsection{Flapping frequency}
We now turn our attention into the dynamical behaviour of the fibres when freely dispersed in the turbulent flow, focusing at first on the main features of their individual motion and deformation. This topic has been investigated by~\citet{rosti2018flexible,rosti2019flowing} in the context of very dilute suspensions, and more recently by~\citet{olivieri2021universal} with the aim of extending the analysis to non-dilute configurations. These previous studies highlighted the potential of using finite-size flexible fibres to measure relevant two-point statistics of turbulence, i.e., longitudinal velocity differences and structure functions, under a proper choice of the fibre's mechanical properties. More specifically, based on the comparison between the characteristic timescales that can be identified in the problem, they showed the existence of two main dynamical regimes, i.e., underdamped or overdamped, differing by the fact that the structural elasticity may manifest or not in the resulting dynamics. Furthermore, the classification could further be divided into different dynamical subregimes. However, only two qualitatively different flapping states were found to ultimately take place~\citep{rosti2019flowing,olivieri2021universal}: (\textit{i}) for sufficiently flexible and/or \added{iso-dense} fibres (i.e., $f_\mathrm{nat}/f_\mathrm{tur} \ll 1$) the dynamics is fully controlled by the flow, and therefore the fibres act effectively as a proxy of the turbulent eddies of comparable size; (\textit{ii}) for sufficiently rigid and inertial fibres (i.e., $f_\mathrm{nat}/f_\mathrm{tur} \gg 1$) the characteristic response time is related to the natural frequency, and the fibre is not dynamically controlled by the turbulent forcing.

In this work, we complement our recent results that generalize such scenario in the case of non-negligible backreaction~\citep{olivieri2021universal}. As a starting point, \cref{fig:flapFreqRatio-vs-NatFreqRatio} shows the behaviour of the \added{flapping frequency $f_\mathrm{flap}$ (identified as the dominant peak frequency from the time history of the fibre's end-to-end distance, as detailed in~\cite{rosti2019flowing})} while varying the ratio between the natural and turbulent eddy frequency. As anticipated, for the \added{denser-than-the-fluid} fibres (corresponding to the underdamped regime) the latter indeed provides a very good indication to distinguish between the two flapping states, which differ from each other for the scaling law of the flapping frequency: for $f_\mathrm{nat}/f_\mathrm{tur} \ll 1$, the flapping frequency is controlled by, and therefore scales with, the turbulent eddy frequency (i.e., $f_\mathrm{flap} = f_\mathrm{tur}$), whereas for $f_\mathrm{nat}/f_\mathrm{tur} \gg 1$ the flapping frequency is given by the natural one (i.e., $f_\mathrm{flap} = f_\mathrm{nat}$). On the other hand, when the fibres are \added{iso-dense} (and thus correspond to the overdamped regime) they are always controlled by turbulence (i.e., $f_\mathrm{flap} = f_\mathrm{tur}$). We remark that the present results concern both dilute and non-dilute cases, with the entity of the backreaction depending on the corresponding mass fraction, as discussed in~\cref{sec:flow-dyn}. The effective qualitative and quantitative characteristics of the modulated flow can thus be appreciably different from the single-phase configuration. To account for this crucial effect, the turbulent eddy frequency is here evaluated as $f_\mathrm{tur} = \alpha' \, \sqrt{S_2} / c$, where $S_2 = \langle (\delta u_\parallel)^2 \rangle$ is the second-order structure function of the longitudinal velocity difference $\delta u_\parallel$ evaluated at the fibre's lengthscale \added{and $\alpha' = 3.0$ is an overall constant  $\mathcal{O}(1)$}. As shown in \cref{fig:flapFreqRatio-vs-NatFreqRatio}, this choice leads to results in overall good agreement with the theoretical prediction even in the non-dilute configurations. Note however, that similar results are found when computing $S_2$ by means of Lagrangian fibre tracking (instead of using directly the fluid flow measured on the Eulerian grid), or if using the average end-to-end distance in place of the fibre length $c$ (with minimal variations that do not appreciably affect the observed scalings). If, on the other hand, one estimated the hydrodynamic frequency using the dimensional arguments based on Kolmogorov scaling in the inertial subrange, a more substantial departure from the reported findings would be obtained for the non-dilute cases where the backreaction is non-negligible~\citep{olivieri2021universal}.
\added{On the other hand, it can be pointed out that the alteration of $f_\mathrm{tur}$ does not have a primary role in the resulting scaling laws (which remain the same already found for the dilute case).}

A final comment  can be made in this regard when considering the (average) elastic energy stored by the fibres during their deformation, shown in the inset of \cref{fig:flapFreqRatio-vs-NatFreqRatio}. \citet{rosti2018flexible} showed that (for a single fibre and in the absence of any appreciable backreaction) this quantity exhibits a maximum when the natural frequency is equal to the turbulence frequency, because of a resonance condition between these two timescales.
 To extend this finding, we choose a representative non-dilute configuration and vary only the bending stiffness in order to substantially retain the same backreaction to the flow (cf.~\cref{tab:caselist_in2}). From the inset of the figure, it clearly appears that, also in the presence of strong turbulence modulation, the average elastic energy $ \langle \mathcal{E}_\mathrm{el} \rangle = \frac{1}{N} \sum_{i=1}^N  \int_0^c \frac{1}{2} \gamma \kappa_i^2 (s) ds$, where $\kappa_i$ is the local bending curvature of the $i$-th fibre, peaks when $f_\mathrm{nat}/f_\mathrm{tur} \approx 1$. \added{Remarkably, the emergence of such peak confirms the idea of a resonance condition in the structural response occurring also in the non-dilute configuration, hence generalizing the theoretical framework previously proposed in the case of negligible backreaction.}

\subsubsection{Maximum curvature}
\begin{figure}
\centering
\includegraphics{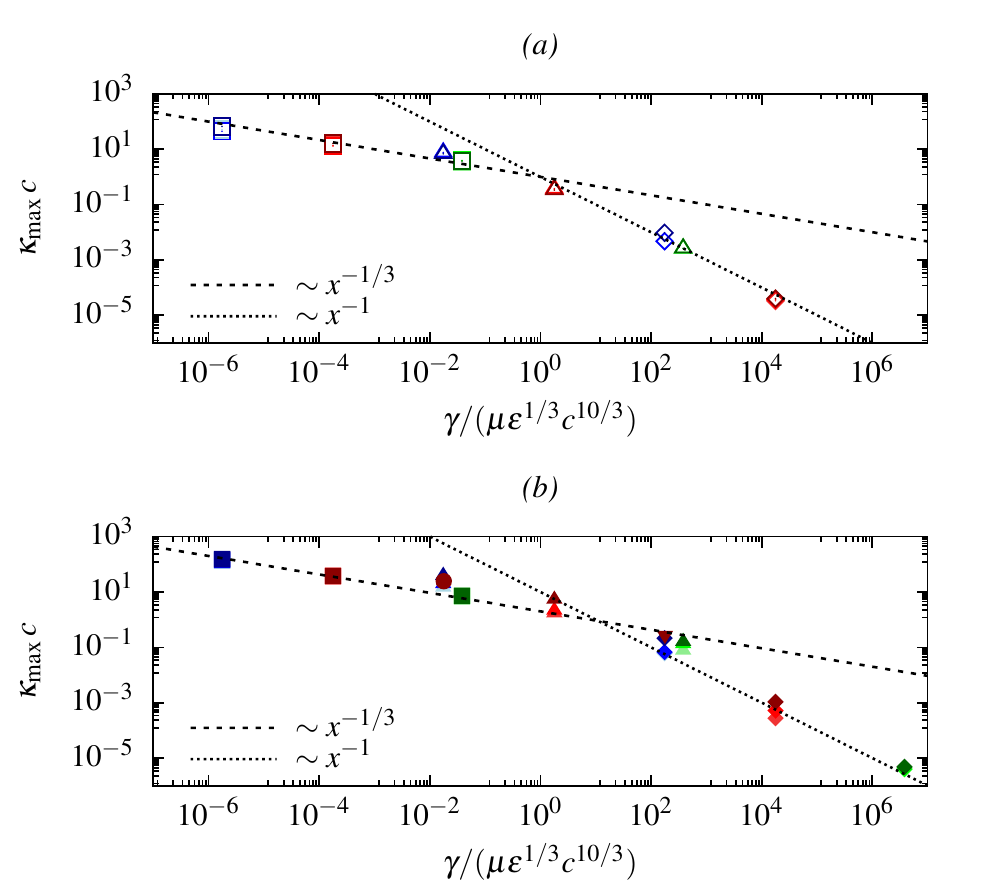}
\caption{Maximum fibre curvature (normalised with the fibre length) as a function of the normalised bending stiffness, for \textit{(a)} \added{iso-dense} and \textit{(b)} \added{denser-than-the-fluid} fibres. The dashed and dotted lines indicate the scaling laws obtained when balancing the forcing term with the inextensibility bending contribution or the linear bending term, respectively. Colors and symbols are used consistently with the indication of \cref{tab:caselist_nb,tab:caselist_in,tab:caselist_in2}.}
\label{fig:Kmax}
\end{figure}

The maximum curvature experienced by the fibres under the action of the flow is another relevant observable when characterising the fibre's deformation, and we report it in \cref{fig:Kmax}. Remarkably, this quantity is of paramount importance in fragmentation processes involving, e.g., fibrous microplastics in oceanic turbulence~\citep{allende2018stretching,brouzet2021laboratory}. Therefore, we focus on understanding how $\kappa_\mathrm{max}$ behaves as a function of the relevant mechanical parameters. We compute the maximum fibre's curvature $\kappa_\mathrm{max}$ as follows: we first evaluate the maximum of $\kappa_i (s)$ for each fibre (at each time instant), and then average this quantity over the different fibres (and over different time instants). While the curvature is expected to be inversely proportional to the fibre's bending stiffness $\gamma$, less trivial is the prediction and corroboration of scaling laws (if any) to qualitatively describe the dependence from the bending stiffness as well as other parameters.

To proceed, we adopt an approach similar to that proposed by~\citet{gay2018characterisation} in the limit case of \added{iso-dense} fibres and dilute suspension, based on balancing different terms in the fibre's dynamical \cref{eq:EB1} estimated by means of dimensional analysis. However, differently from~\citet{gay2018characterisation}, here we directly compare these terms without introducing an effective elastic length nor focusing on an energy balance, and consequently obtain different scaling laws for the maximum curvature.  For the sake of simplicity, the same assumptions, i.e., one-way coupling and inertial subrange scaling, are retained (and will be commented on later). For the forcing term we can write $\Fb \sim \mu \, (\epsilon c)^{1/3}$, where $\mu$ is the dynamic viscosity, while the (linear) bending term can be dimensionally expressed as $\gamma \partial^4_s \Xb \sim  \gamma \kappa_\mathrm{max} /c^2$. Moreover, as shown by~\citet{marheineke2006fiber,gay2018characterisation}, from the tension term $\partial_s \left( T \partial_s \Xb \right)$ one can isolate the contribution $\partial_s \left( \gamma \kappa^2 \partial_s \Xb \right)$ ensuring the inextensibility constraint under bending deformations, which can be further split in two terms, one parallel to the fibre's axis and one parallel to the curvature vector; focusing solely on the latter (cubic) contribution, we have $ \partial_s \left( T \partial_s \Xb \right) \sim \gamma \kappa_\mathrm{max}^3$. Remarkably,  the cubic term is subleading with respect to the linear one for $\kappa_\mathrm{max} c \ll 1$, while the opposite occurs for $\kappa_\mathrm{max} c \gg 1$. Therefore, we can predict the following two regimes:
(\textit{i}) for sufficiently rigid fibres, the maximum curvature is given by the balance between the forcing and the linear bending term, thus obtaining that $\kappa_\mathrm{max} c \sim \left[ \gamma / (\mu \epsilon^{1/3} c^{10/3})\right]^{-1}$;
(\textit{ii}) for sufficiently flexible fibres, the balance instead is between the forcing and the cubic bending contribution to the tension term, yielding $\kappa_\mathrm{max} c \sim \left[ \gamma / (\mu \epsilon^{1/3} c^{10/3})\right]^{-1/3}$.

\Cref{fig:Kmax} shows the results from our numerical simulations along with the expected scaling laws, both for the \added{iso-dense} (top) and \added{denser-than-the-fluid} fibre cases (bottom). Overall, we observe a good agreement of the numerical results with respect to both proposed scalings, with only a limited deviation for the \added{denser-than-the-fluid} fibres approximately in the transition region between the two regimes. Nevertheless, some remarks on the underlying assumptions are needed to properly outline the limitations of the proposed model. First, using the inertial subrange scaling for the turbulence forcing is strictly justified only for sufficiently dilute suspensions and fibres of intermediate length. Second, although we note from the numerical results a systematic increase of $\kappa_\mathrm{max}$ with the fibre's inertia (i.e., linear density difference), the influence of the latter is not accounted for in the derivation of the scaling laws. Consequently, it was not possible to obtain a unique master curve collecting all the data together. 
We note the difference with the scalings that were obtained for the flapping frequency, where the inertial term was indeed considered and had a crucial role to distinguish between the overdamped and underdamped cases. Here instead the same qualitative behaviour is obtained for both the \added{iso-dense} and \added{denser-than-the-fluid} fibres.
Nevertheless, our predictions for the remaining set of parameters are corroborated not only for \added{iso-dense}, but also for the \added{denser-than-the-fluid} inertial fibres, thus being general for the widest parametric range.

\subsubsection{\added{Clustering}}
\begin{figure}
\centering
\includegraphics{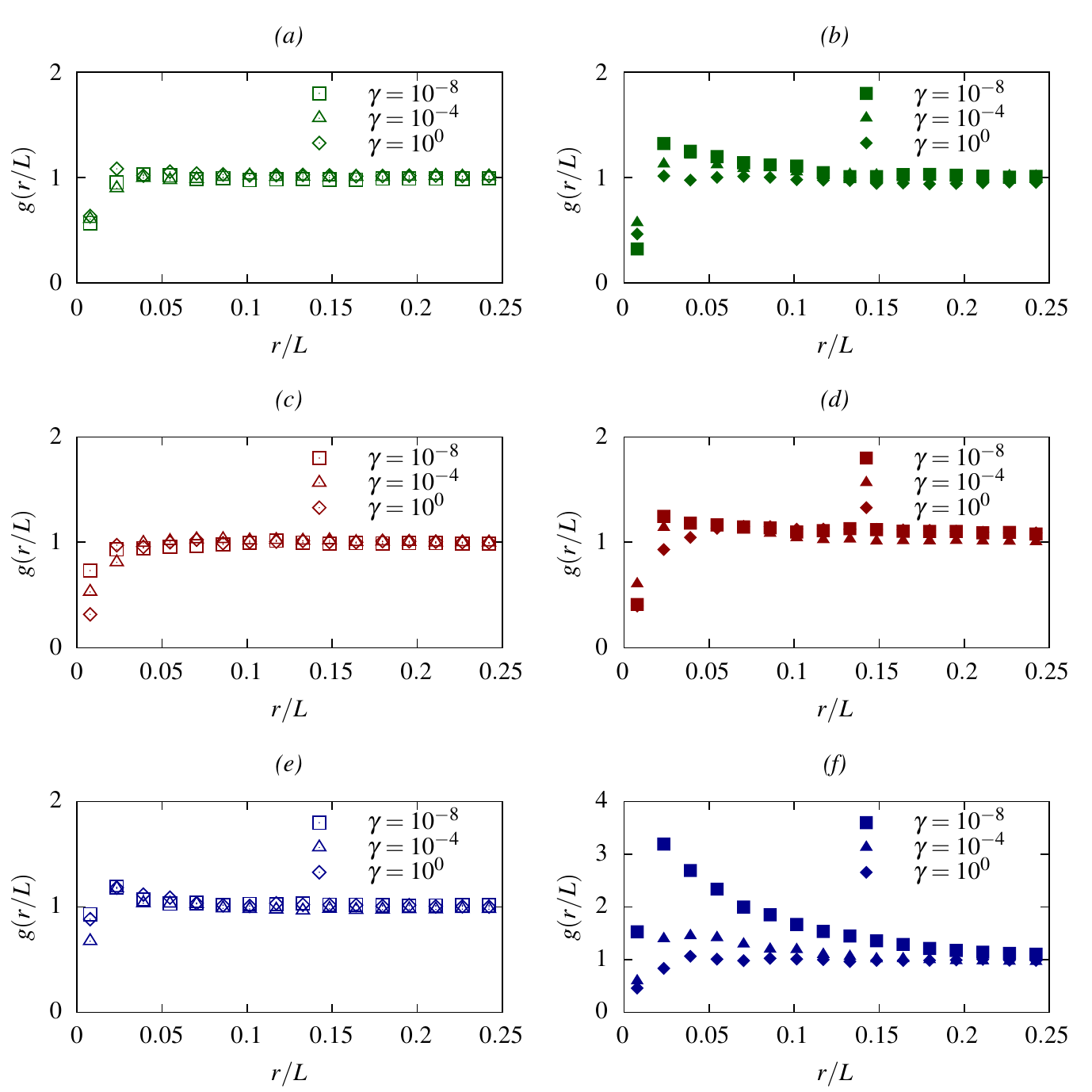}
\caption{\added{Radial distribution function for cases with $N=10^3$ \added{iso-dense} (left) or \added{denser-than-the-fluid} (right) fibres of different length (top: short, intermediate: centre, long: bottom). Colors and symbols are used consistently with the indication of \cref{tab:caselist_nb,tab:caselist_in}.}}
\label{fig:RDF_all}
\end{figure}

After the characterisation of the individual motion of the dispersed fibres, we consider some relevant aspects concerning their collective dynamics. We look in particular at the local concentration of the suspension to inspect the presence of clustering phenomena, a key feature in the analysis of turbulent particle-laden flows~\citep{eaton1994preferential,bec2007heavy} and with a specific relevance in the context of fibres aggregation~\citep{lundell2011fluid,verhille2017structure}. Different metrics have been proposed to quantify the tendency of particles to preferentially accumulate along with sampling specific regions of the flow~\citep{brandt2021particle}, among which is the radial distribution (or pair correlation) function (RDF)~\citep{saw2008inertial,salazar2008experimental,olivieri2014effect}. Such quantity indicates the probability of having a pair of particles at a given mutual distance, thus highlighting the presence of accumulation at particular lengthscales, while assuming a constant unit value when the particle distribution is locally uniform. Here, we compute the RDF using its classical definition for point-like particles and specifically considering the Lagrangian midpoints, being concerned with the mutual distance between the different anisotropic particles. Furthermore, we restrict the analysis to the cases with the largest number of fibres $N=10^3$ to ensure well-converged statistics.

\Cref{fig:RDF_all} shows the RDF obtained from the DNS results, from which several observations can be made. The left panels (\cref{fig:RDF_all}\textit{a,c,e}) refer to the \added{iso-dense} fibres of different length and bending stiffness, for which the clustering effect is always essentially negligible. Indeed, such evidence is consistent with the crucial role of inertia in \added{the formation of clustering}~\citep{eaton1994preferential}. On the other hand, the cases for \added{denser-than-the-fluid} fibres in the right panels (\cref{fig:RDF_all}\textit{b,d,f}) reveal a richer phenomenology. First, for all the investigated fibre lengths, it turns out that the accumulation increases when decreasing the bending stiffness. This systematic trend can be explained by the fact that, when the fibres are more flexible, they can adapt more easily to the local flow structure and thus sample more frequently the zones of lower vorticity similarly to what observed for point-like particles. Conversely, the fibres are prevented to do so when increasing $\gamma$, because of the rigidity constraint, and consequently a weaker effect is found. 
As another prominent effect, we observe that the highest peak in the RDF is found for the long fibres (\cref{fig:RDF_all}\textit{f}), whereas less remarkable variations are observed for the short and intermediate ones (\cref{fig:RDF_all}\textit{b,d}). 
\added{Such finding can be explained indeed in terms of the combined role of inertia (i.e., Stokes number) and flexibility (i.e., effective compliance), such that for the longest fibres clustering is more favoured.
  }

\begin{figure}
\centering
\includegraphics{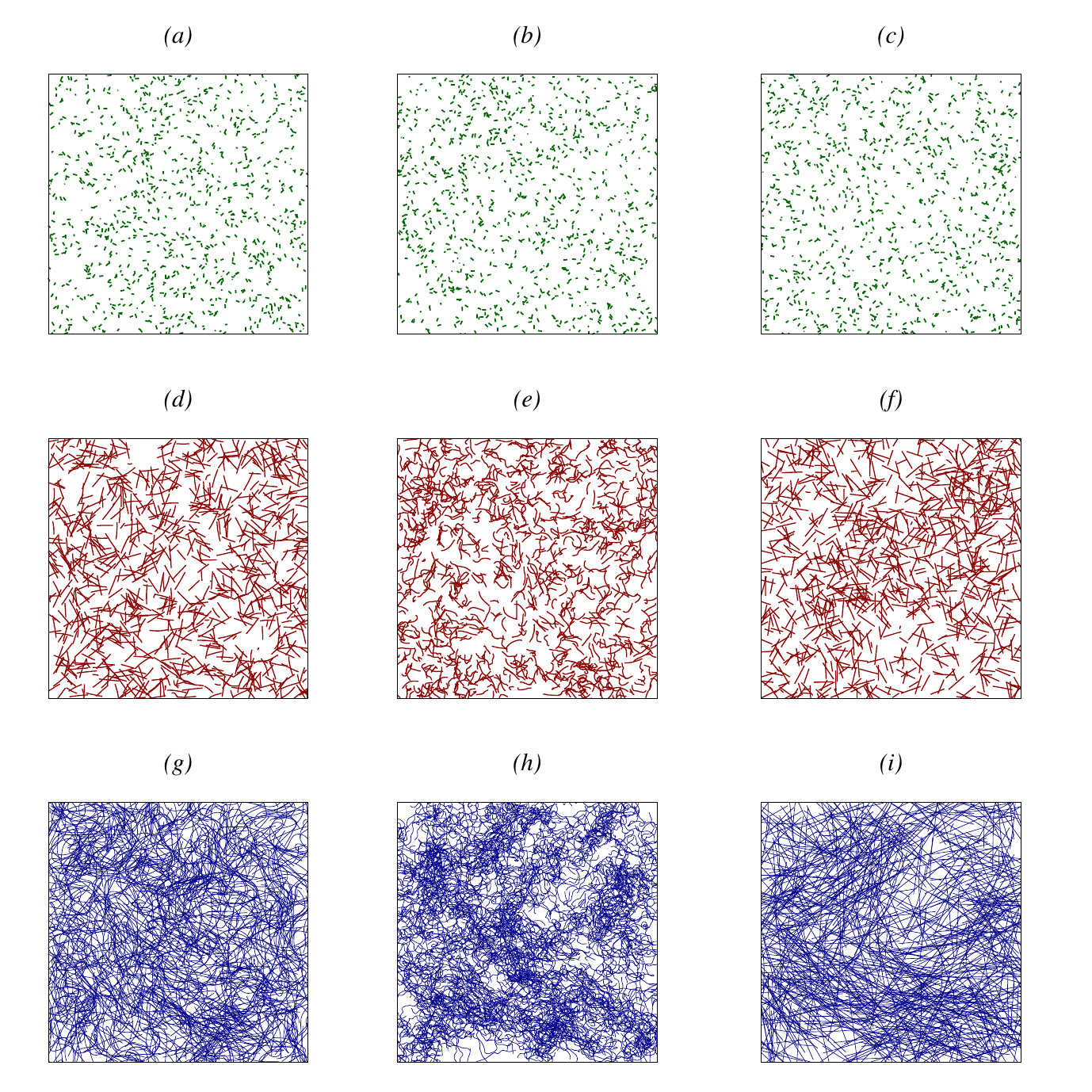}
\caption{Two-dimensional snapshots of fibres of different length (top: short; middle: intermediate; bottom: long) and for three different configurations (left: \added{iso-dense} with intermediate bending stiffness; center: \added{denser-than-the-fluid} with lowest bending stiffness; right: \added{denser-than-the-fluid} with highest bending stiffness).}
\label{fig:clusteringVis_3x3}
\end{figure}

\Cref{fig:clusteringVis_3x3} provides a qualitative insight by collecting some instantaneous visualisations for the \added{iso-dense} cases of intermediate stiffness (left panels), and for both the most flexible (centre) and most rigid (right) \added{denser-than-the-fluid} fibres. For \added{iso-dense} fibres clustering is indeed always negligible, without any peculiar role observed for the fibre's flexibility. Conversely, for \added{denser-than-the-fluid} fibres \added{the influence of the latter becomes stronger} as the length is increased. As a result, for long fibres a remarkable difference can be noted between the flexible and rigid case (\cref{fig:clusteringVis_3x3}\textit{h,i}), with much more intense clusters that can detected for the former.

\added{
As a final comment, we highlight that clustering does not appear to have a clear role in the turbulence modulation: as shown in \cref{sec:flow-dyn}, it is the mass fraction the most relevant (and arguably unique) control parameter, without a specific effect found, e.g., for the fibre’s length, in the backreaction mechanism. 
}

%As a final comment, we note that to properly isolate the peculiar mechanisms driving the preferential concentration for such finite-size and anisotropic objects requires further evidence that should be obtained by properly varying the parameters governing this particular phenomenon. Indeed, along with the aforementioned influence of the bending stiffness, the preferential concentration is expected to be controlled by the interplay between the inertia of the dispersed objects and suitable characteristic timescales of the flow (i.e., analogously to the Stokes number for point-like particles); at the same time, the (local) backreaction to the carrier flow can also have a non-negligible effect on the entity of clustering, thus significantly increasing the complexity of the problem. Furthermore, the dependence with respect to the length of the objects and the role of inextensibility should be also investigated. This represents a challenging task that is left for future work.

\subsubsection{Preferential alignment}
As the final step of our investigation, we focus on a peculiar issue of anisotropic particles in turbulence~\citep{voth2017anisotropic}: the statistical characterisation of the alignment between the fibre orientation and some specific quantities of the turbulent flow (i.e., vorticity and principal directions of the strain rate). The topic has been extensively investigated in the framework of fibres with infinitesimal length, and more recently for finite-size fibres~\citep{pumir2011orientation,ni2015measurements,pujara2019scale,pujara2021shape}, pointing out the preferential alignment of sufficiently short fibres with the fluid vorticity and that of longer fibres with the most extensional eigenvector of the strain rate. We note, however, that such theoretical and computational studies still often rely on a one-way coupling assumption and are typically based on the classical Jeffery's model (the latter strictly holding only for fibres shorter than the dissipative lengthscale). Moreover, fibres are usually assumed to be \added{iso-dense} and rigid. Here, we tackle the problem using a \added{four-way coupled} simulation approach and span the wide parametric range in terms of fibre's inertia, length and flexibility.

\begin{figure}
\centering
\includegraphics{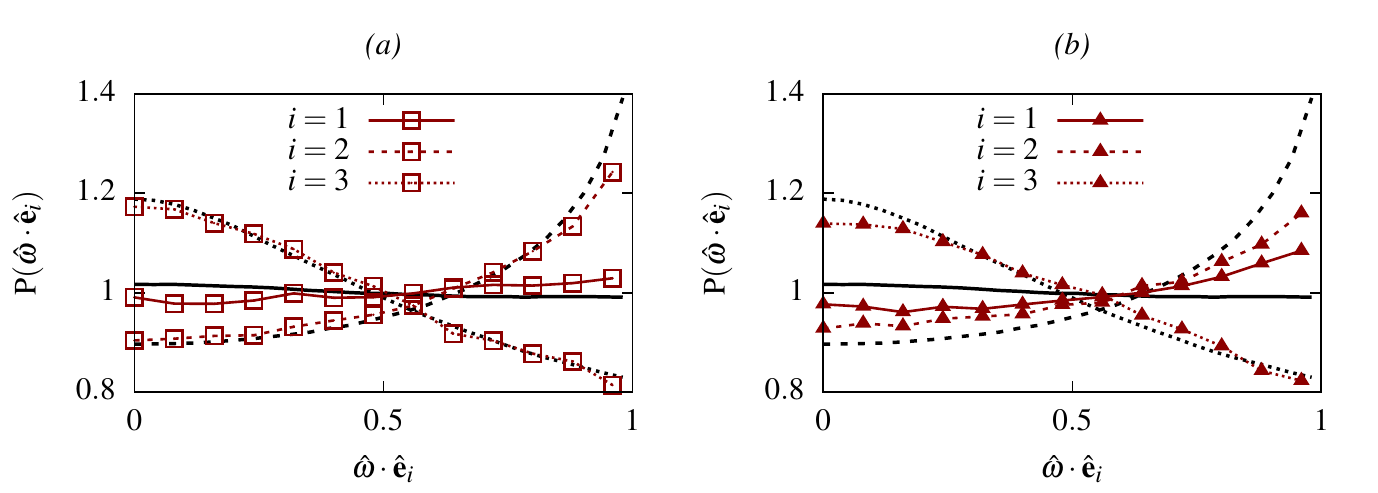}
\caption{PDFs of the alignment between the fluid vorticity and eigenvectors of the strain rate computed in the fully Eulerian framework for the single-phase case (black lines without symbols) and via a Lagrangian coarse-grained approach (brown lines with symbols) for two representative cases \added{(at intermediate length)}: \textit{(a)} \added{iso-dense} fibres \added{(with lowest bending stiffness)}, \textit{(b)} \added{denser-than-the-fluid} fibres \added{(with intermediate bending stiffness).}}
\label{fig:alignment_0}
\end{figure}

\added{We compare the fibre's local orientation (i.e., considering each segment connecting two adjacent Lagrangian points, and then averaging the results over different segments and fibres) with the local fluid flow} to identify the existence of preferential alignment, and to explore how the latter varies with the main properties of the suspension. Since in our four-way coupled model the fluid flow is locally perturbed by the presence of the fibre, when computing the vorticity and strain rate we employ a coarse-graining procedure to overcome such effect in the proximity of the Lagrangian points. Specifically, a stencil of $7^3$ Eulerian cells surrounding the Lagrangian point is used. To assess the validity of this choice, at first we employ this procedure to compute the alignment between the vorticity and strain rate, a well-known feature of homogeneous turbulent flows~\citep{ashurst1987alignment,tsinober1992experimental}. Results obtained in a representative configuration with \added{iso-dense} fibres (for which the backreaction is overall negligible) are shown in \cref{fig:alignment_0}\textit{a}, along with those evaluated using the fully Eulerian framework in the single-phase case: the same qualitative outcome is observed between the Lagrangian coarse-grained and fully Eulerian approaches, with the vorticity resulting mostly aligned with the intermediate eigenvector and orthogonal to the direction of maximum compression, i.e., the third eigenvector~\citep{tsinober1992experimental}.
\begin{figure}
\centering
\includegraphics{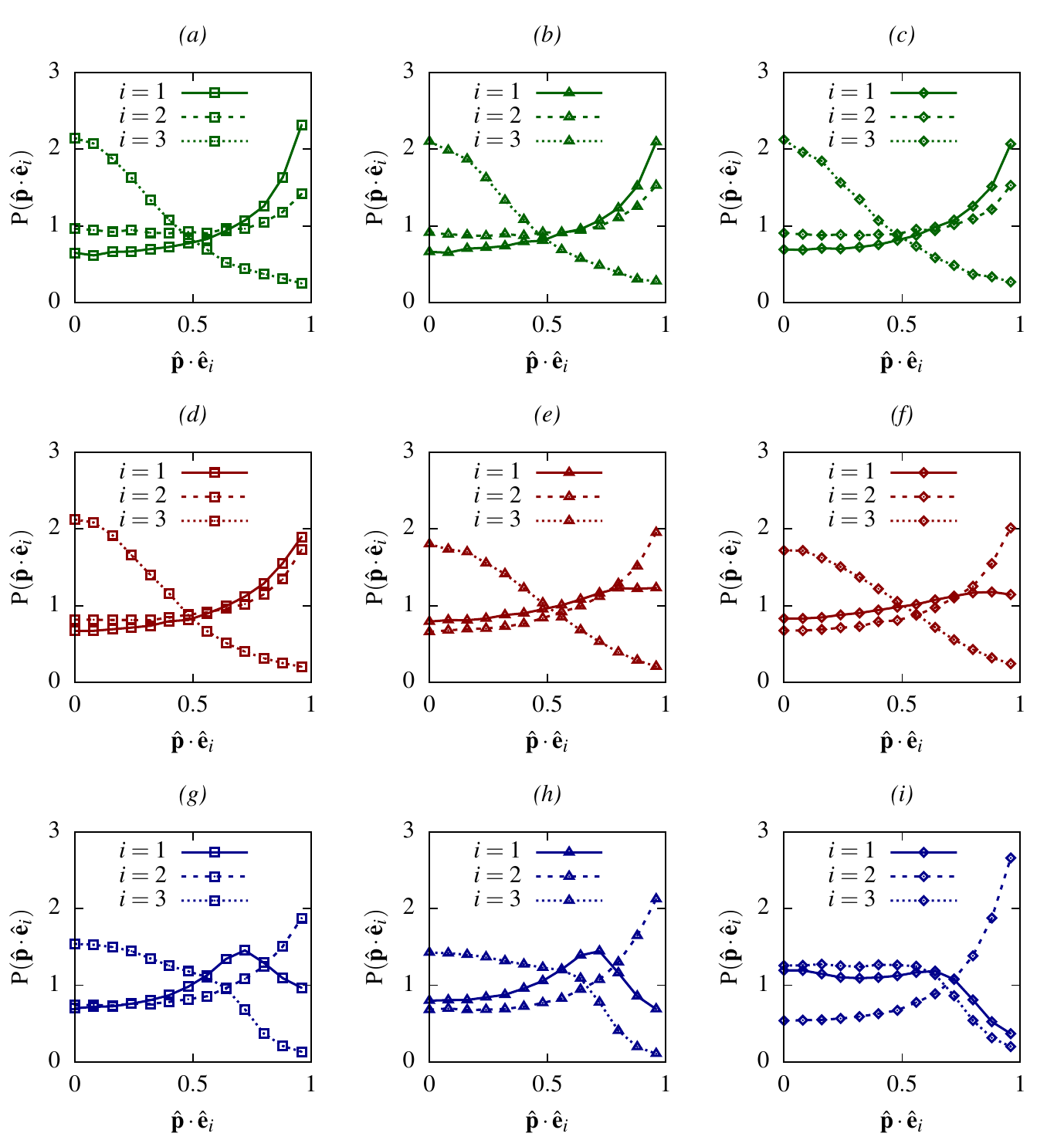}
\caption{PDFs of the alignment between the (local) fibre's orientation and the eigenvectors of the strain rate, for \added{iso-dense} fibres of different length (top: short; middle: intermediate; bottom: long) and bending stiffness (increasing from left to right panels). The solid, dashed and dotted lines indicate the alignment with the first, second and third eigenvector of the strain rate, respectively.  Colors and symbols are used consistently with the indication of \cref{tab:caselist_nb,tab:caselist_in}.}
\label{fig:alignment_EigS-DX_neutr}
\end{figure}
On the other hand, only a minimal departure may be noted for the alignment with the first eigenvector (corresponding to the direction of maximum extension). Nevertheless, in light of the overall agreement the coarse-grained approach can be safely exploited for the following analysis.  Furthermore, the same quantities are shown in \cref{fig:alignment_0}\textit{b} for a case with \added{denser-than-the-fluid} fibres which cause a significant modulation of the turbulent flow. In this case, the alteration in terms of vorticity-strain alignment appears not dramatic yet not completely negligible. Specifically, the main difference that can be observed is the mild attenuation of the alignment of vorticity with the intermediate eigenvector. Note that, this effect can play indirectly a role when observing the alignment of inertial fibres with the flow.

Let us therefore consider the alignment between the fibre's orientation and the strain rate principal directions, starting from the \added{iso-dense} cases reported in~\cref{fig:alignment_EigS-DX_neutr}. In particular, we first consider the results for short fibres (\cref{fig:alignment_EigS-DX_neutr}\textit{a,b,c}) for which, regardless of the variation in the bending stiffness, a strong alignment with both the first and second eigenvectors is found, along with anti-alignment with the third one. Remarkably, this finding is in very good agreement with the experimental results reported by~\citet{ni2015measurements} (see Fig. 8 therein), specifically for what it concerns the predominant alignment with the first (rather than the second) eigenvector (conversely, as noted by~\citet{ni2015measurements} such qualitative agreement was not found when using the one-way coupling approach). We note the resemblance in the parameters (i.e., fibre length and density, micro-scale Reynolds number) between these cases and the experimental study by~\citet{ni2015measurements}; in particular, here the fibre is longer, yet still comparable, than the Kolmogorov lengthscale. If, on the other hand, we increase the fibre length to the intermediate value (\cref{fig:alignment_EigS-DX_neutr}\textit{d,e,f}), we observe that the same phenomenology is retained only for the most flexible case, while the alignment with the most extensional direction is progressively lost when increasing the bending stiffness. As a result, for essentially rigid fibres of intermediate length (i.e., belonging to the inertial subrange), the dominant alignment is found with the intermediate eigenvector. Finally, when the length is further increased so that the fibres are comparable with the integral lengthscale (\cref{fig:alignment_EigS-DX_neutr}\textit{g,h,i}), several qualitative modifications are observed: the strongest alignment is always found with the intermediate eigenvector (further increasing with the rigidity of the fibres), while the PDF relative to the third eigenvector gets more flattened (both compared with the shorter lengths and when increasing the bending stiffness). Furthermore, for flexible fibres the PDF relative to the first eigenvector shows now a peak at an intermediate value. Such non-monotonic behaviour, which is not observed for other configurations, appears to be conditioned by having a relatively low bending stiffness, and a possible transition can be argued towards a more anti-aligned situation when further increasing the bending stiffness.

\begin{figure}
\centering
\includegraphics{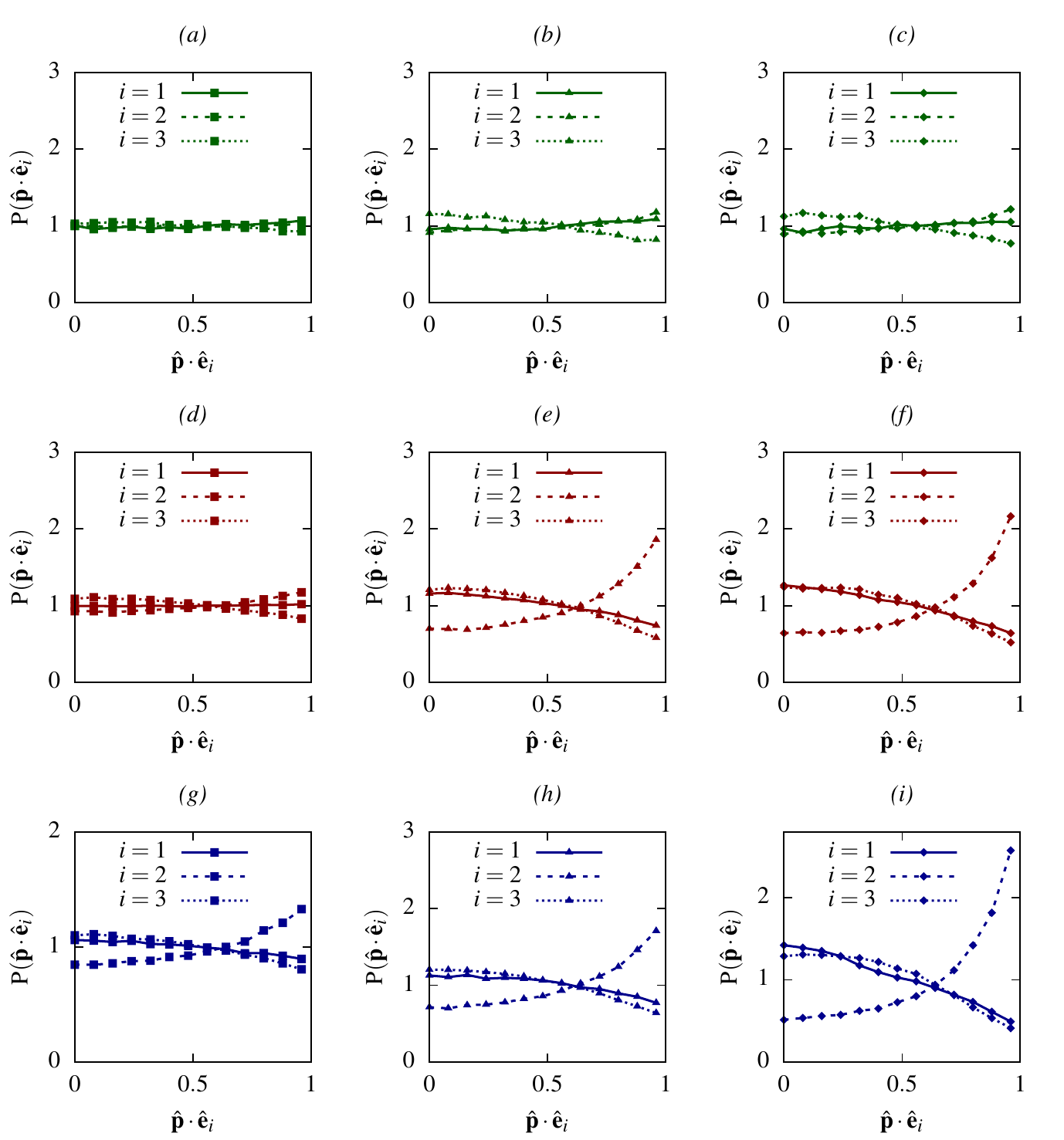}
\caption{PDFs of the alignment between the (local) fibre's orientation and the eigenvectors of the strain rate, for \added{denser-than-the-fluid} fibres of different length (top: short; middle: intermediate; bottom: long) and bending stiffness (increasing from left to right panels). The solid, dashed and dotted lines indicate the alignment with the first, second and third eigenvector of the strain rate, respectively.  Colors and symbols are used consistently with the indication of \cref{tab:caselist_nb,tab:caselist_in}.}
\label{fig:alignment_EigS-DX_heavy}
\end{figure}

\Cref{fig:alignment_EigS-DX_heavy} shows the results concerning the same configurations but for \added{denser-than-the-fluid} fibres. The resulting scenario is now dramatically different compared with the \added{iso-dense} cases, with two main effects that can be noticed. First, with respect to the \added{iso-dense} cases, the preferential alignment (or anti-alignment) of fibres with the strain rate turns out to be substantially weaker. This is especially evident when considering short fibres (\cref{fig:alignment_EigS-DX_heavy}\textit{a,b,c}) or more flexible ones (\cref{fig:alignment_EigS-DX_heavy}\textit{d,g}), where all curves are much more horizontal, with an approximately constant value around $1$, thus indicating a tendency towards the situation of random alignment. \added{On one hand, this result could be intuitively expected since the dynamics for highly inertial particles is significantly delayed with respect to the forcing acted by the turbulent flow.
As a result, at a given instant the local fibre's orientation and surrounding flow topology are substantially less correlated with a departure from the preferential alignment.}
 However, when increasing the rigidity (i.e., bending stiffness) the situation systematically changes: for all the considered fibre lengths, we observe a stronger alignment with the intermediate eigenvector, along with a more pronounced anti-alignment with the third eigenvector. A difference can be noted instead in terms of the alignment with the first eigenvector between short fibres, which  show a weak alignment, and intermediate or long fibres, for which a moderate anti-alignment is noticed. Overall, it can be underlined that the inertia of the fibres and the rigidity constraint play an apparently opposite role, i.e., depleting vs promoting the preferential alignment of the dispersed particles with the flow. %Finally, we remark that, despite these findings (obtained with a fully-resolved approach) looked in good agreement when compared with available experimental evidence in a specific configuration, more refined studies are definitely deserved in order to better characterise and understand the main features of preferential alignment for finite-size, inertial and flexible fibres.

\section{Conclusions and outlook}
\label{sec:conclusions}
In this work, we have investigated the mutually coupled dynamics of dispersed fibres in homogeneous isotropic turbulence by means of \added{four-way coupled} DNS. The interaction between finite-size fibres and the turbulent flow has been tackled in the framework of an immersed boundary method to properly model the coupling between the two phases, thus going beyond the typical assumptions of infinitesimal length and one-way coupling. Fixing a reference single-phase condition, such methodology has been employed to perform an extensive numerical study \added{(in zero-gravity condition)} over the main parameters that control the dynamics of both the dispersed and carrier phase, i.e., the fibre's linear density, length and bending stiffness, as well as the number of fibres, in order to simulate and compare a variety of configurations (i.e., \added{iso-dense} vs \added{denser-than-the-fluid}, short vs long, flexible vs rigid, and dilute vs non-dilute). Several outcomes can be outlined from our numerical results, concerning both the modulation of the turbulent flow and some relevant features of the suspension, such as the fibre's flapping states and deformation, as well as the possibility of clustering and preferential orientation.

To start, we have systematically described how the backreaction of the dispersed phase affects the macroscopic behaviour of the carrier flow, with an overall depletion of the turbulent kinetic energy that is reflected by the remarkable variation of the (decreasing) micro-scale Reynolds number. In agreement with reported evidence for other particle-laden flows, such macroscopic effect is controlled by the mass fraction of the suspension, and not by other reference quantities such as the number density or volume fraction.  Indeed, we remark that the backreaction effect can be non-negligible already at number densities that correspond to the so-called dilute regime~\citep{butler2018microstructural}. The mass fraction appears to be the main control parameter also when analysing the modification of the energy spectra and the corresponding scale-by-scale energy transfer. When increasing the mass fraction, two robust features are observed from the energy spectra, i.e., an overall large-scale energy depletion along with a relative increase of the small-scale energy content. Such evidence can be better explained by looking at the energy fluxes in Fourier space, where a general tendency is observed as the mass fraction is increased, with the suppression of the nonlinear term along with the increasing relevance of the energy transfer directly associated with the fluid-solid coupling. Finally, we point out that the fibre's bending stiffness appears to have a minor yet systematic influence, with a weaker backreaction for more flexible fibres, as it could be intuitively argued.

Regarding the fibre dynamics, we have first characterised the fibre's deformation in terms of the dominant flapping frequency and possible flapping states, revisiting the phenomenological model of~\citet{rosti2018flexible,rosti2019flowing} for the case of non-negligible turbulence modulation. As a result, the same qualitative scenario is observed, with only two possible dynamical regimes: \textit{(i)} for relatively flexible fibres (such that their natural frequency is lower than the hydrodynamic frequency of turbulence eddies of comparable size) or overdamped (such as \added{iso-dense}) fibres, the dynamics is fully controlled by the turbulence structures at the fibre's lengthscale, i.e., the flapping frequency is locked to that of turbulence, with potential application in exploiting the fibre to measure the two-point statistics of the flow;  \textit{(ii)} for relatively rigid (and inertial) fibres, the fibre's flapping frequency is conversely decoupled from the flow and manifesting the natural structural response to the fluid forcing. From a similar perspective, we have characterised the maximum curvature that fibres can experience while being transported and deformed by the carrier flow, outlining the existence of two different scaling laws whose physical explanation is in partial agreement with that proposed by a previous study dealing with \added{iso-dense} fibres~\citep{gay2018characterisation}.
Next, we have looked at the local concentration of the dispersed fibres in order to detect the occurrence of clustering phenomena. Following an approach similar to that usually employed for particle-laden flows, based on evaluating the radial distribution function over pairs of fibres, we observed a systematic increase of \added{clustering} when increasing the fibre's flexibility (i.e., decreasing the bending stiffness) and, as expected, for inertial fibres. In particular, we found the strongest \added{clustering} for the longest and most flexible, \added{denser-than-the-fluid} fibres. %Nonetheless, understanding the key parameters controlling such effect remains an open issue that is left for future investigations.
Finally, we have assessed the preferential alignment between the fibre's (local) orientation and the strain rate principal directions, comparing our findings with those reported by previous studies in similar conditions, as well as presenting novel results for the flexible or inertial cases. 
Overall, we observe that both the alignment with the first or second principal direction (i.e., of maximum extension or intermediate extension/compression) and the anti-alignment with the third principal direction (i.e., of maximum compression) typically increase when the fibres become more rigid, whereas remarkably decrease when moving from \added{iso-dense} to \added{denser-than-the-fluid} fibres. 

In conclusion, we outline possible developments of the present study that could provide the motivation for future work: on one hand, exploring the behaviour of finite-size fibre suspensions at substantially higher Reynolds numbers, in particular to assess whether the same or more complex turbulence modulation mechanisms are occurring. On the other hand, further efforts are deserved to better understand the specific features of this kind of suspensions in terms of \added{clustering} and, even more importantly, preferential alignment.
\added{For inertial fibres, another remarkable point is exploring how the dynamics is altered in finite Froude number conditions (i.e., when the gravitional forces are not negligible).
Lastly, another framework deserving further research efforts is surely represented by wall-bounded turbulent flows, which are beyond the scope of the present investigation.}

\backsection[Acknowledgements]{S.O. and M.E.R. acknowledge the computational resources provided on the Deigo cluster by the Scientific Computing and Data Analysis section of Research Support Division at OIST and on the Fugaku supercomputer by RIKEN through the HPCI System Research Project (Project IDs: hp210229 and hp210269).}

\backsection[Funding]{The research was supported by the Okinawa Institute of Science and Technology Graduate University (OIST) with subsidy funding from the Cabinet Office, Government of Japan. A.M. acknowledges the financial support from the Compagnia di San Paolo, Project MINIERA No. I34I20000380007, and from the Project PoC – BUYT MAIH, No. C36I20000140006.
}

\backsection[Declaration of interests]{The authors report no conflict of interest.}

\backsection[Data availability statement]{The data that support the findings of this study are available from the corresponding authors upon reasonable request.}

\backsection[Author ORCIDs]{\\
S. Olivieri, \href{https://orcid.org/0000-0002-7795-6620}{https://orcid.org/0000-0002-7795-6620};\\
A. Mazzino, \href{https://orcid.org/0000-0003-0170-2891}{https://orcid.org/0000-0003-0170-2891};\\
M. E. Rosti, \href{https://orcid.org/0000-0002-9004-2292}{https://orcid.org/0000-0002-9004-2292}.}

\appendix

\section{\added{Evaluation of collision events}}
\label{app:collisions}
\begin{figure}
    \centering
    \includegraphics{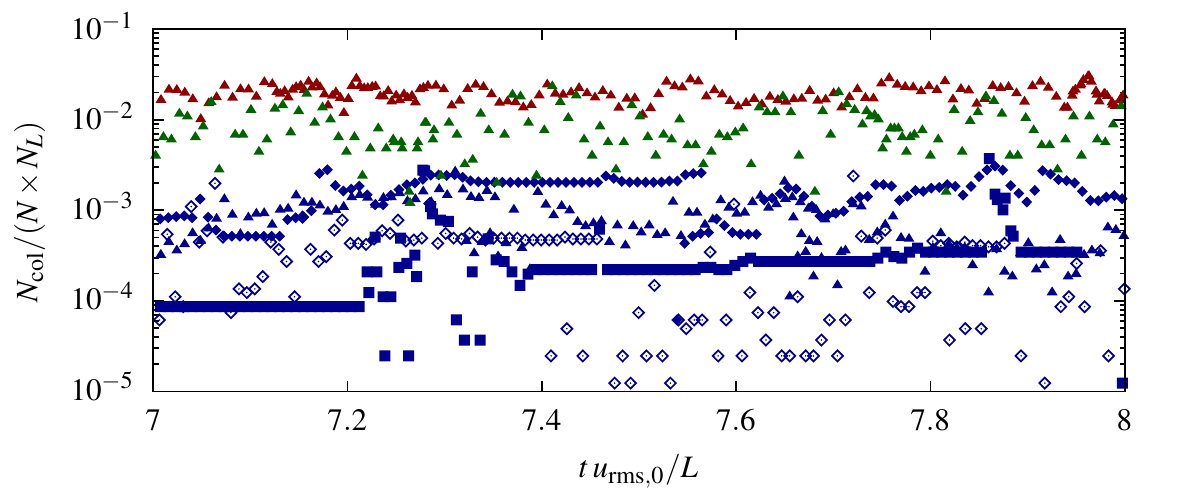}
    \caption{\added{Normalised number of collision over time for several representative cases. Colors and symbols are used consistently with the indication of \cref{tab:caselist_nb,tab:caselist_in}.}}
    \label{fig:collisionsNorm}
\end{figure}

\added{To account for the effective relevance of the collision model in the performed simulations, we have monitored over time the number of collision events (occurring when two Lagrangian points belonging to different fibres are found below the critical distance $\Delta_\mathrm{col}$). To properly estimate the relative importance of the fiber-to-fiber interactions, we normalize such number with the total number of Lagrangian point. In~\cref{fig:collisionsNorm}, such quantity is shown for cases with different fibre length, density and bending stiffness. Overall, it can be noticed that the occurrence of collisions is always very limited, including cases with the highest concentration, such events typically involving less than $1\%$ of the ensemble of fibre material points.
}

\begin{figure}
    \centering
    \includegraphics{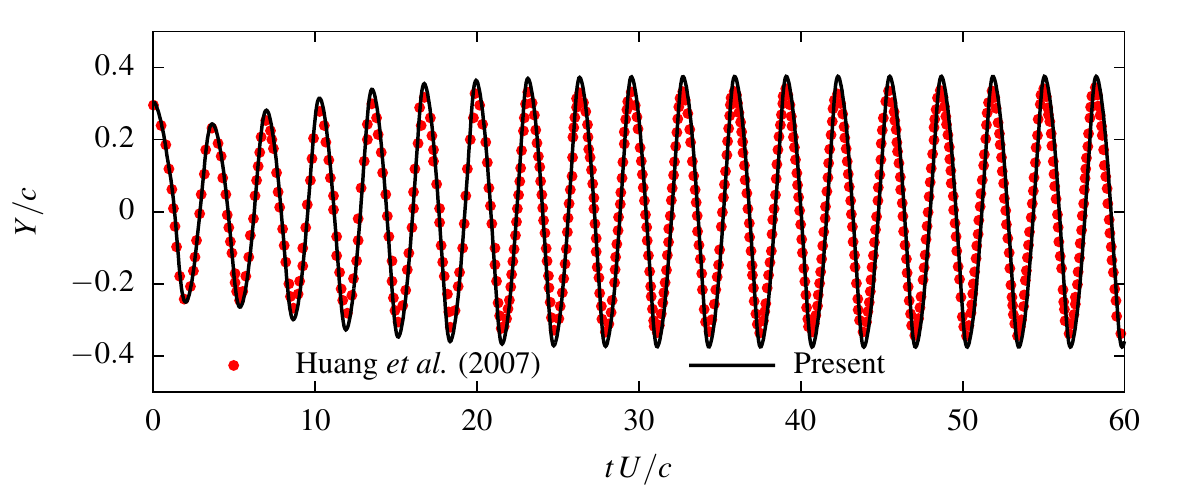}
    \caption{\added{Benchmark results} for the case of a flapping filament in uniform flow, showing the time history of the filament's trailing point position by~\citet{huang_shin_sung_2007a} (red circles) and that obtained with the code used for the present study (black line).}
    \label{fig:valHuang}
\end{figure}

\section{\added{Benchmark for the numerical technique}}
\label{sec:validation}
The numerical code used in the present analysis has been extensively validated in the past. For the sake of completeness, in this appendix we present the result of an additional test case.  In particular, we consider a passively flapping filament in a uniform flow. For a description of the case setup the reader is referred to~\citet{huang_shin_sung_2007a} and, in particular, Fig. 13 therein. Fig.~\ref{fig:valHuang} shows the comparison between our results and those by the original paper, where one can note the good agreement both in the amplitude and frequency of the flapping oscillation. Such evidence, therefore, further confirms the \added{validity} of our numerical findings.

\section{List of configurations investigated in the parametric study}
\label{sec:list}
This appendix provides additional information on the setup chosen for our investigation and, in particular, the combinations of parameters that have been considered, which are listed in \cref{tab:caselist_nb,tab:caselist_in,tab:caselist_in2}. Note that, the parameters $\Delta \widetilde{\rho}$, $c$ and $\gamma$ are given in code units since in the case of finite-size fibres it is not straightforward to choose a representative nondimensional form. The information is therefore complemented by that regarding the carrier flow: the kinematic viscosity is set to $\nu = 1.59 \times 10^{-3}$ and the fluid density is set to 1, while the parameters of the external forcing are such that the resulting turbulence is characterised in the single-phase case by a r.m.s. of the velocity fluctuations $u_\mathrm{rms} \approx 5.14 \times 10^{-1}$ and average dissipation rate $\epsilon \approx 4.5 \times 10^{-2}$, and consequently in the Kolmogorov dissipative lengthscale $\eta = (\nu^3/\epsilon)^{1/4} \approx1.7\times 10^{-2}$ and the micro-scale Reynolds number $\mathrm{Re}_\lambda \approx 120$.

\Cref{tab:caselist_nb} reports the cases with almost \added{iso-dense} fibres (indicated with empty symbols) and \cref{tab:caselist_in} the cases with \added{iso-dense} fibres (denoted instead with filled symbols); for the former, the resulting solid-to-fluid density ratio (which can only be estimated since the fibre's diameter is not explicitly controlled) is around 1 while for the latter it is approximately 500. Additionally, for a particular configuration, i.e., \added{iso-dense} fibres with intermediate length and density, a refined study was performed over the bending stiffness, as detailed in \cref{tab:caselist_in2}.
To give a further overall indication, we can compare the three considered fibre lengths with the dissipative lengthscale and the domain length (used as a rough estimate for the integral lenghtscale), yielding $c/\eta = \left\{ 6, 30, 116 \right\}$ and $c/L = \left\{ 1.6\times10^{-2}, 8.0\times10^{-2}, 0.3 \right\}$, respectively.

\definecolor{light-green}{RGB}{128,238,124}
\definecolor{light-red}{RGB}{234,26,39}
\definecolor{light-blue}{RGB}{160,208,223}
\definecolor{dark-green}{RGB}{10,83,1}
\definecolor{dark-red}{RGB}{120,0,1}
\definecolor{dark-blue}{RGB}{0,0,122}

\begin{table}
\centering
%\begin{ruledtabular}
\begin{tabular}{cccccccc}
%\begin{tabular}{clllllll}
% \toprule
\rule{0pt}{3ex} \rule[-1.2ex]{0pt}{0pt}Marker & $\Delta \widetilde{\rho} $ & $c$ & $\gamma$ & $N$ & $n\,c^3$ & $\phi_V$ & $\mathcal{M}$ \\
%\colrule
 {\tikz{\node[draw,scale=0.4,regular polygon, regular polygon sides=4, color=light-green](){};}} & $10^{-3}$ & 0.1 & $10^{-8}$ & $10^1$ & $1.6\times 10^{-4}$  & $ 3.0\times 10^{-5}$ & $ 1.2\times 10^{-5}$ \\
 {\tikz{\node[draw,scale=0.4,regular polygon, regular polygon sides=4, color=green        ](){};}} & $10^{-3}$ & 0.1 & $10^{-8}$ & $10^2$ & $1.6\times 10^{-3}$  & $ 3.0\times 10^{-4}$ & $ 1.2\times 10^{-4}$ \\
 {\tikz{\node[draw,scale=0.4,regular polygon, regular polygon sides=4, color=dark-green](){};}} & $10^{-3}$ & 0.1 & $10^{-8}$ & $10^3$ & $1.6\times 10^{-2}$  & $ 3.0\times 10^{-3}$ & $ 1.2\times 10^{-3}$ \\
 {\tikz{\node[draw,scale=0.3,regular polygon, regular polygon sides=3, color=light-green](){};}} & $10^{-3}$ & 0.1 & $10^{-4}$ & $10^1$ & $1.6\times 10^{-4}$  & $ 3.0\times 10^{-5}$ & $ 1.2\times 10^{-5}$  \\
 {\tikz{\node[draw,scale=0.3,regular polygon, regular polygon sides=3, color=green        ](){};}} & $10^{-3}$ & 0.1 & $10^{-4}$ & $10^2$ & $1.6\times 10^{-3}$  & $ 3.0\times 10^{-4}$ & $ 1.2\times 10^{-4}$  \\
 {\tikz{\node[draw,scale=0.3,regular polygon, regular polygon sides=3, color=dark-green](){};}} & $10^{-3}$ & 0.1 & $10^{-4}$ & $10^3$ & $1.6\times 10^{-2}$  & $ 3.0\times 10^{-3}$ & $ 1.2\times 10^{-3}$   \\
 {\tikz{\node[draw,scale=0.4,regular polygon, regular polygon sides=4, rotate=45, color=light-green](){};}} & $10^{-3}$ & 0.1 & $10^{0}$ & $10^1$ & $1.6\times 10^{-4}$  & $ 3.0\times 10^{-5}$ & $ 1.2\times 10^{-5}$    \\
 {\tikz{\node[draw,scale=0.4,regular polygon, regular polygon sides=4, rotate=45, color=green        ](){};}} & $10^{-3}$ & 0.1 & $10^{0}$ & $10^2$ & $1.6\times 10^{-3}$  & $ 3.0\times 10^{-4}$ & $ 1.2\times 10^{-4}$   \\
 {\tikz{\node[draw,scale=0.4,regular polygon, regular polygon sides=4, rotate=45, color=dark-green](){};}} & $10^{-3}$ & 0.1 & $10^{0}$ & $10^3$ & $1.6\times 10^{-2}$  & $ 3.0\times 10^{-3}$ & $ 1.2\times 10^{-3}$   \\ 
 {\tikz{\node[draw,scale=0.4,regular polygon, regular polygon sides=4, color=light-red](){};}} & $10^{-3}$ & 0.5 & $10^{-8}$ & $10^1$  &  $2.0\times 10^{-2}$  & $ 1.5\times 10^{-4}$ & $ 5.8\times 10^{-5}$ \\
 {\tikz{\node[draw,scale=0.4,regular polygon, regular polygon sides=4, color=red        ](){};}} & $10^{-3}$ & 0.5 & $10^{-8}$ & $10^2$  &  $2.0\times 10^{-1}$ &  $ 1.5\times 10^{-3}$ & $ 5.8\times 10^{-4}$ \\
 {\tikz{\node[draw,scale=0.4,regular polygon, regular polygon sides=4, color=dark-red](){};}} & $10^{-3}$ & 0.5 & $10^{-8}$ & $10^3$  &  $2.0\times 10^{0}$ &  $ 1.5\times 10^{-2}$ & $ 5.8\times 10^{-3}$ \\
 {\tikz{\node[draw,scale=0.3,regular polygon, regular polygon sides=3, color=light-red](){};}} & $10^{-3}$ & 0.5 & $10^{-4}$ & $10^1$  &  $2.0\times 10^{-2}$  & $ 1.5\times 10^{-4}$ & $ 5.8\times 10^{-5}$  \\
 {\tikz{\node[draw,scale=0.3,regular polygon, regular polygon sides=3, color=red      ](){};}} & $10^{-3}$ & 0.5 & $10^{-4}$ & $10^2$   &  $2.0\times 10^{-1}$ &  $ 1.5\times 10^{-3}$ & $ 5.8\times 10^{-4}$ \\
 {\tikz{\node[draw,scale=0.3,regular polygon, regular polygon sides=3, color=dark-red](){};}} & $10^{-3}$ & 0.5 & $10^{-4}$ & $10^3$  &  $2.0\times 10^{0}$ &  $ 1.5\times 10^{-2}$ & $ 5.8\times 10^{-3}$  \\
 {\tikz{\node[draw,scale=0.4,regular polygon, regular polygon sides=4, rotate=45, color=light-red](){};}} & $10^{-3}$ & 2.0 & $10^{0}$ & $10^1$  &  $2.0\times 10^{-2}$  & $ 1.5\times 10^{-4}$ & $ 5.8\times 10^{-5}$  \\
 {\tikz{\node[draw,scale=0.4,regular polygon, regular polygon sides=4, rotate=45, color=red        ](){};}} & $10^{-3}$ & 2.0 & $10^{0}$ & $10^2$  &  $2.0\times 10^{-1}$ &  $ 1.5\times 10^{-3}$ & $ 5.8\times 10^{-4}$  \\
 {\tikz{\node[draw,scale=0.4,regular polygon, regular polygon sides=4, rotate=45, color=dark-red](){};}} & $10^{-3}$ & 2.0 & $10^{0}$ & $10^3$  &  $2.0\times 10^{0}$ &  $ 1.5\times 10^{-2}$ & $ 5.8\times 10^{-3}$  \\ 
 {\tikz{\node[draw,scale=0.4,regular polygon, regular polygon sides=4, color=light-blue](){};}} & $10^{-3}$ & 2.0 & $10^{-8}$ & $10^1$  &  $1.3\times 10^{0}$  & $ 6.1\times 10^{-4}$ & $ 2.3\times 10^{-4}$ \\
 {\tikz{\node[draw,scale=0.4,regular polygon, regular polygon sides=4, color=blue        ](){};}} & $10^{-3}$ & 2.0 & $10^{-8}$ & $10^2$  &  $1.3\times 10^{1}$  & $ 6.1\times 10^{-3}$ & $ 2.3\times 10^{-3}$ \\
 {\tikz{\node[draw,scale=0.4,regular polygon, regular polygon sides=4, color=dark-blue](){};}} & $10^{-3}$ & 2.0 & $10^{-8}$ & $10^3$  &  $1.3\times 10^{2}$  & $ 6.1\times 10^{-2}$ & $ 2.3\times 10^{-2}$ \\
 {\tikz{\node[draw,scale=0.3,regular polygon, regular polygon sides=3, color=light-blue](){};}} & $10^{-3}$ & 2.0 & $10^{-4}$ & $10^1$  &  $1.3\times 10^{0}$  & $ 6.1\times 10^{-4}$ & $ 2.3\times 10^{-4}$ \\ 
 {\tikz{\node[draw,scale=0.3,regular polygon, regular polygon sides=3, color=blue      ](){};}} & $10^{-3}$ & 2.0 & $10^{-4}$ & $10^2$  &  $1.3\times 10^{1}$  & $ 6.1\times 10^{-3}$ & $ 2.3\times 10^{-3}$  \\
 {\tikz{\node[draw,scale=0.3,regular polygon, regular polygon sides=3, color=dark-blue](){};}} & $10^{-3}$ & 2.0 & $10^{-4}$ & $10^3$  &  $1.3\times 10^{2}$  & $ 6.1\times 10^{-2}$ & $ 2.3\times 10^{-2}$  \\
 {\tikz{\node[draw,scale=0.4,regular polygon, regular polygon sides=4, rotate=45, color=light-blue](){};}} & $10^{-3}$ & 2.0 & $10^{0}$ & $10^1$ &  $1.3\times 10^{0}$  & $ 6.1\times 10^{-4}$ & $ 2.3\times 10^{-4}$ \\ 
 {\tikz{\node[draw,scale=0.4,regular polygon, regular polygon sides=4, rotate=45, color=blue        ](){};}} & $10^{-3}$ & 2.0 & $10^{0}$ & $10^2$ &  $1.3\times 10^{1}$  & $ 6.1\times 10^{-3}$ & $ 2.3\times 10^{-3}$   \\
 {\tikz{\node[draw,scale=0.4,regular polygon, regular polygon sides=4, rotate=45, color=dark-blue](){};}} & $10^{-3}$ & 2.0 & $10^{0}$ & $10^3$  &  $1.3\times 10^{2}$  & $ 6.1\times 10^{-2}$ & $ 2.3\times 10^{-2}$  \\ 
%\botrule
\end{tabular}
\caption{List of parametric combinations considered in the present work (where $\Delta \widetilde{\rho}$ is the linear density difference between the fibre and the fluid, $c$ is the fibre's length, $\gamma$ is the fibre's bending stiffness, $N$ is the number of fibres dispersed in the triperiodic domain box; all values are given in code units). The corresponding number density $n\,c^3$, volume fraction $\phi_V$ and mass fraction $\mathcal{M}$ are also reported. Each case is denoted by the marker used in the figures throughout the paper. The present table shows the cases for \added{iso-dense} fibres, whereas \added{iso-dense} fibres are reported in \cref{tab:caselist_in}. }
\label{tab:caselist_nb}
%\end{ruledtabular}
\end{table}

\begin{table}
\centering
%\begin{ruledtabular}
\begin{tabular}{cccccccc}
% \toprule
\rule{0pt}{3ex} \rule[-1.2ex]{0pt}{0pt}Marker & $\Delta \widetilde{\rho} $ & $c$ & $\gamma$ & $N$ & $n\,c^3$ & $\phi_V$ & $\mathcal{M}$ \\
%\colrule
 {\tikz{\node[scale=0.4,regular polygon, regular polygon sides=4, fill=light-green](){};}} & $10^{0}$ & 0.1 & $10^{-8}$ & $10^1$ & $1.6\times 10^{-4}$  & $ 3.0\times 10^{-5}$ & $ 4.0\times 10^{-3}$   \\
 {\tikz{\node[scale=0.4,regular polygon, regular polygon sides=4, fill=green        ](){};}} & $10^{0}$ & 0.1 & $10^{-8}$ & $10^2$ & $1.6\times 10^{-3}$  & $ 3.0\times 10^{-4}$ & $ 3.8\times 10^{-2}$   \\
 {\tikz{\node[scale=0.4,regular polygon, regular polygon sides=4, fill=dark-green](){};}} & $10^{0}$ & 0.1 & $10^{-8}$ & $10^3$ & $1.6\times 10^{-2}$  & $ 3.0\times 10^{-3}$ & $ 2.9\times 10^{-1}$   \\
 {\tikz{\node[scale=0.3,regular polygon, regular polygon sides=3, fill=light-green](){};}} & $10^{0}$ & 0.1 & $10^{-4}$ & $10^1$ & $1.6\times 10^{-4}$  & $ 3.0\times 10^{-5}$ & $ 4.0\times 10^{-3}$   \\
 {\tikz{\node[scale=0.3,regular polygon, regular polygon sides=3, fill=green        ](){};}} & $10^{0}$ & 0.1 & $10^{-4}$ & $10^2$ & $1.6\times 10^{-3}$  & $ 3.0\times 10^{-4}$ & $ 3.8\times 10^{-2}$   \\
 {\tikz{\node[scale=0.3,regular polygon, regular polygon sides=3, fill=dark-green](){};}} & $10^{0}$ & 0.1 & $10^{-4}$ & $10^3$ & $1.6\times 10^{-2}$  & $ 3.0\times 10^{-3}$ & $ 2.9\times 10^{-1}$   \\
 {\tikz{\node[scale=0.4,regular polygon, regular polygon sides=4, rotate=45, fill=light-green](){};}} & $10^{0}$ & 0.1 & $10^{0}$ & $10^1$  & $1.6\times 10^{-4}$  & $ 3.0\times 10^{-5}$ & $ 4.0\times 10^{-3}$  \\
 {\tikz{\node[scale=0.4,regular polygon, regular polygon sides=4, rotate=45, fill=green        ](){};}} & $10^{0}$ & 0.1 & $10^{0}$ & $10^2$  & $1.6\times 10^{-3}$  & $ 3.0\times 10^{-4}$ & $ 3.8\times 10^{-2}$  \\
 {\tikz{\node[scale=0.4,regular polygon, regular polygon sides=4, rotate=45, fill=dark-green](){};}} & $10^{0}$ & 0.1 & $10^{0}$ & $10^3$  & $1.6\times 10^{-2}$  & $ 3.0\times 10^{-3}$ & $ 2.9\times 10^{-1}$  \\ 
 {\tikz{\node[scale=0.4,regular polygon, regular polygon sides=4, fill=light-red](){};}} & $10^{0}$ & 0.5 & $10^{-8}$ & $10^1$ &  $2.0\times 10^{-2}$  & $ 1.5\times 10^{-4}$ & $ 2.0\times 10^{-2}$   \\
 {\tikz{\node[scale=0.4,regular polygon, regular polygon sides=4, fill=red        ](){};}} & $10^{0}$ & 0.5 & $10^{-8}$ & $10^2$ &  $2.0\times 10^{-1}$ &  $ 1.5\times 10^{-3}$ & $ 1.7\times 10^{-1}$   \\
 {\tikz{\node[scale=0.4,regular polygon, regular polygon sides=4, fill=dark-red](){};}} & $10^{0}$ & 0.5 & $10^{-8}$ & $10^3$ &  $2.0\times 10^{0}$ &  $ 1.5\times 10^{-2}$ & $ 6.7\times 10^{-1}$   \\
 {\tikz{\node[scale=0.3,regular polygon, regular polygon sides=3, fill=light-red](){};}} & $10^{0}$ & 0.5 & $10^{-4}$ & $10^1$  &  $2.0\times 10^{-2}$  & $ 1.5\times 10^{-4}$ & $ 2.0\times 10^{-2}$  \\
 {\tikz{\node[scale=0.3,regular polygon, regular polygon sides=3, fill=red      ](){};}} & $10^{0}$ & 0.5 & $10^{-4}$ & $10^2$  &  $2.0\times 10^{-1}$ &  $ 1.5\times 10^{-3}$ & $ 1.7\times 10^{-1}$  \\
 {\tikz{\node[scale=0.3,regular polygon, regular polygon sides=3, fill=dark-red](){};}} & $10^{0}$ & 0.5 & $10^{-4}$ & $10^3$ &  $2.0\times 10^{0}$ &  $ 1.5\times 10^{-2}$ & $ 6.7\times 10^{-1}$   \\
 {\tikz{\node[scale=0.4,regular polygon, regular polygon sides=4, rotate=45, fill=light-red](){};}} & $10^{0}$ & 2.0 & $10^{0}$ & $10^1$ &  $2.0\times 10^{-2}$  & $ 1.5\times 10^{-4}$ & $ 2.0\times 10^{-2}$   \\
 {\tikz{\node[scale=0.4,regular polygon, regular polygon sides=4, rotate=45, fill=red        ](){};}} & $10^{0}$ & 2.0 & $10^{0}$ & $10^2$ &  $2.0\times 10^{-1}$ &  $ 1.5\times 10^{-3}$ & $ 1.7\times 10^{-1}$   \\
 {\tikz{\node[scale=0.4,regular polygon, regular polygon sides=4, rotate=45, fill=dark-red](){};}} & $10^{0}$ & 2.0 & $10^{0}$ & $10^3$  &  $2.0\times 10^{0}$ &  $ 1.5\times 10^{-2}$ & $ 6.7\times 10^{-1}$  \\ 
 {\tikz{\node[scale=0.4,regular polygon, regular polygon sides=4, fill=light-blue](){};}} & $10^{0}$ & 2.0 & $10^{-8}$ & $10^1$ &  $1.3\times 10^{0}$  & $ 6.1\times 10^{-4}$ & $ 7.5\times 10^{-2}$    \\
 {\tikz{\node[scale=0.4,regular polygon, regular polygon sides=4, fill=blue        ](){};}} & $10^{0}$ & 2.0 & $10^{-8}$ & $10^2$ &  $1.3\times 10^{1}$  & $ 6.1\times 10^{-3}$ & $ 4.5\times 10^{-1}$   \\
 {\tikz{\node[scale=0.4,regular polygon, regular polygon sides=4, fill=dark-blue](){};}} & $10^{0}$ & 2.0 & $10^{-8}$ & $10^3$ &  $1.3\times 10^{2}$  & $ 6.1\times 10^{-2}$ & $ 8.9\times 10^{-1}$   \\
 {\tikz{\node[scale=0.3,regular polygon, regular polygon sides=3, fill=light-blue](){};}} & $10^{0}$ & 2.0 & $10^{-4}$ & $10^1$  &  $1.3\times 10^{0}$  & $ 6.1\times 10^{-4}$ & $ 7.5\times 10^{-2}$  \\
 {\tikz{\node[scale=0.3,regular polygon, regular polygon sides=3, fill=blue      ](){};}} & $10^{0}$ & 2.0 & $10^{-4}$ & $10^2$    &   $1.3\times 10^{1}$  & $ 6.1\times 10^{-3}$ & $ 4.5\times 10^{-1}$  \\
 {\tikz{\node[scale=0.3,regular polygon, regular polygon sides=3, fill=dark-blue](){};}} & $10^{0}$ & 2.0 & $10^{-4}$ & $10^3$  &  $1.3\times 10^{2}$  & $ 6.1\times 10^{-2}$ & $ 8.9\times 10^{-1}$  \\
 {\tikz{\node[scale=0.4,regular polygon, regular polygon sides=4, rotate=45, fill=light-blue](){};}} & $10^{0}$ & 2.0 & $10^{0}$ & $10^1$  &  $1.3\times 10^{0}$  & $ 6.1\times 10^{-4}$ & $ 7.5\times 10^{-2}$   \\
 {\tikz{\node[scale=0.4,regular polygon, regular polygon sides=4, rotate=45, fill=blue        ](){};}} & $10^{0}$ & 2.0 & $10^{0}$ & $10^2$  &  $1.3\times 10^{1}$  & $ 6.1\times 10^{-3}$ & $ 4.5\times 10^{-1}$  \\
 {\tikz{\node[scale=0.4,regular polygon, regular polygon sides=4, rotate=45, fill=dark-blue](){};}} & $10^{0}$ & 2.0 & $10^{0}$ & $10^3$  &  $1.3\times 10^{2}$  & $ 6.1\times 10^{-2}$ & $ 8.9\times 10^{-1}$  \\ 
%\botrule
\end{tabular}
\caption{Same as \cref{tab:caselist_nb}, but referring to the cases with \added{iso-dense} fibres.}
\label{tab:caselist_in}
%\end{ruledtabular}
\end{table}

\begin{table}
\centering
%\begin{ruledtabular}
\begin{tabular}{cccccccc}
% \toprule
\rule{0pt}{3ex} \rule[-1.2ex]{0pt}{0pt}Marker & $\Delta \widetilde{\rho} $ & $c$ & $\gamma$ & $N$ & $n\,c^3$ & $\phi_V$ & $\mathcal{M}$ \\
%\colrule 
 {\tikz{\node[scale=0.4,regular polygon, regular polygon sides=4, fill=dark-red](){};}} & $10^{0}$ & 0.5 & $10^{-8}$ & $10^3$ &  $2.0\times 10^{0}$ &  $ 1.5\times 10^{-2}$ & $ 6.7\times 10^{-1}$   \\
 {\tikz{\node[scale=0.4, circle, fill=dark-red](){};}} & $10^{0}$ & 0.5 & $10^{-6}$ & $10^3$  &  $2.0\times 10^{0}$ &  $ 1.5\times 10^{-2}$ & $ 6.7\times 10^{-1}$   \\
 {\tikz{\node[scale=0.3,regular polygon, regular polygon sides=3, fill=dark-red](){};}} & $10^{0}$ & 0.5 & $10^{-4}$ & $10^3$  &  $2.0\times 10^{0}$ &  $ 1.5\times 10^{-2}$ & $ 6.7\times 10^{-1}$   \\
 {\tikz{\node[scale=0.3,regular polygon, regular polygon sides=3, rotate=180, fill=dark-red](){};}} & $10^{0}$ & 0.5 & $10^{-2}$ & $10^3$  &  $2.0\times 10^{0}$ &  $ 1.5\times 10^{-2}$ & $ 6.7\times 10^{-1}$   \\
 {\tikz{\node[scale=0.4,regular polygon, regular polygon sides=4, rotate=45, fill=dark-red](){};}} & $10^{0}$ & 0.5 & $10^{0}$ & $10^3$ &  $2.0\times 10^{0}$ &  $ 1.5\times 10^{-2}$ & $ 6.7\times 10^{-1}$   \\
%\botrule
\end{tabular}
\caption{Refined study over $\gamma$ for \added{iso-dense} fibres.}
\label{tab:caselist_in2}
%\end{ruledtabular}
\end{table}

\section{Derivation of the scale-by-scale energy transfer}
\label{app:spectral-balance} 
In this appendix, we briefly show how to derive the scale-by-scale energy transfer discussed in \cref{sec:scale-by-scale}, along with identifying the various terms appearing in such balance. For a more detailed explanation, the reader is referred to classical textbooks on turbulence theory, e.g.,~\citet{pope2000turbulent}.

As the starting point, we perform the Fourier transform of~\cref{eq:NS1,eq:NS2},
\begin{equation}
  \partial_t \hat{\ub} + \hat{\mathbf{G}} = - i \mathbf{k} \hat{p}/\rho_\mathrm{f} - \nu k^2 \hat{\ub} + \hat{\fb}_\mathrm{tur} + \hat{\fb}_\mathrm{fs},
  \label{eq:NS1_spectr}
\end{equation}
\begin{equation}
  \mathbf{k} \cdot \hat{\ub} = 0,
  \label{eq:NS2_spectr}    
\end{equation}
where $\hat{(\cdot)}(\mathbf{k},t) = \mathcal{F}\{ (\cdot)(\xb,t) \}$ denotes the Fourier transform from physical to spectral space, $\mathbf{k}$ is the wavenumber vector and $k$ its magnitude, $i$ is the imaginary unit and $\mathbf{G}$ represents the nonlinear term appearing in the momentum equation. Additionally, the same equations can be written for the complex conjugate $\hat{\ub}^*$.

 We now multiply~\cref{eq:NS1_spectr} by $\hat{\ub}^*$, so that the pressure term drops due to the incompressibility constraint. The same applies in the momentum equation for $\hat{\ub}^*$ when multiplying by $\hat{\ub}$. Then, we sum the two equations for $\hat{\ub}$ and $\hat{\ub}^*$, thus obtaining an equation for the spectral kinetic energy $\hat{E}(\mathbf{k},t) \equiv  \langle \hat{\ub}^* \cdot \hat{\ub} \rangle / 2$ which reads
\begin{equation}
  \partial_t \hat{E} = \hat{T} + \hat{V} + \hat{F}_\mathrm{tur} + \hat{F}_\mathrm{fs}.
  \label{eq:spectral_energy}
\end{equation}
Four different contributions can be identified in the equation above:
\begin{itemize}
    \item[] $\hat{T} = - \frac{1}{2} \, (\hat{\mathbf{G}} \cdot \hat{\ub}^* + \hat{\mathbf{G}}^* \cdot \hat{\ub}) $ is the energy transfer associated with the nonlinear term;
    \item[] $ \hat{V} = - 2 \nu k^2 \hat{E} $ is the energy dissipation associated with the viscous term;
    \item[] $\hat{F}_\mathrm{tur} = \frac{1}{2} \, (\hat{\fb}_\mathrm{tur} \cdot \hat{\ub}^* + \hat{\fb}^*_\mathrm{tur} \cdot \hat{\ub})$ is the energy input associated with the turbulence forcing;
    \item[] $\hat{F}_\mathrm{fs} = \frac{1}{2} \, (\hat{\fb}_\mathrm{fs} \cdot \hat{\ub}^* + \hat{\fb}^*_\mathrm{fs} \cdot \hat{\ub})$ is the energy transfer associated with the fluid-structure coupling.
\end{itemize}

Next, by isotropically averaging~\cref{eq:spectral_energy} over the generic sphere of radius $k$, we obtain a similar equation for the energy spectrum $E(k,t)$,
\begin{equation}
  \partial_t {E} = {T} + {V} + {F}_\mathrm{tur} + {F}_\mathrm{fs}.
\end{equation}
Assuming a statistically stationary state, the left-hand-side vanishes. To obtain the scale-by-scale energy transfer, we first integrate \cref{eq:spectral_energy} from $k$ to infinity, yielding
\begin{equation}
  0 = \Pi + D' + P + \Pi_\mathrm{fs},
\end{equation}
 where the various contributions appearing in the balance are
\begin{equation}
\Pi(k) = \int_k^\infty T (k) \, \mathrm{d}k,
\end{equation}
\begin{equation}
D'(k) = \int_k^\infty V(k) \, \mathrm{d}k,
\end{equation}
\begin{equation}
P(k) = \int_k^\infty F_\mathrm{tur}(k) \, \mathrm{d}k,
\end{equation}
\begin{equation}
\Pi_\mathrm{fs}(k) = \int_k^\infty F_\mathrm{fs}(k) \, \mathrm{d}k.
\end{equation}
Again, these quantities are associated with the nonlinear, dissipative, external forcing and fluid-solid coupling terms, respectively. Remarkably, not only the nonlinear transfer $\Pi(k)$, but also the transfer associated with the fluid-solid coupling $\Pi_\mathrm{fs} (k)$ are energy fluxes, without any net energy input/output, i.e., $\Pi(0) = \Pi_\mathrm{fs}(0) = 0$, with the overall balance governed only by the external forcing and viscous dissipation, i.e.,
\begin{equation}
\int_0^\infty F_\mathrm{tur}(k) \, \mathrm{d}k = - \int_0^\infty V(k) \, \mathrm{d}k = \epsilon.
\end{equation}
Indeed, exploiting this result we recast the definition of the dissipation using $D(k) = - \int_0^k V(k) \, \mathrm{d}k = \epsilon + D'(k)$ in place of $D'(k)$, and we finally obtain \cref{eq:spectral-power-balance}. Note that, this choice is meant only for better describing the scale-by-scale energy transfer when plotting the various terms. 

%\bibliographystyle{jfm}
%%\bibliographystyle{abbrvnat}
%\bibliography{references}

\end{document}